\documentclass[%
 reprint,
 amsmath,amssymb,
 aps,
]{revtex4-2}

\usepackage{graphicx}
\usepackage{dcolumn}
\usepackage{bm}
\usepackage{subcaption}
\usepackage{hyperref}
\usepackage{booktabs}
\usepackage{ragged2e}
\usepackage{tabularx}
\usepackage[all]{nowidow}

\begin{document}

\preprint{APS/123-QED}

\title{Analysis of electron spectra dynamics in a moving periodical ponderomotive potential}

\author{Marek Kuchař}
 \email{marek.kuchar@mff.cuni.cz}
\author{Kamila Moriová}
\author{Martin Kozák}%
\affiliation{%
 Department of Chemical Physics and Optics, Faculty of Mathematics and Physics,
Charles University, Ke Karlovu 3, Prague CZ-12116, Czech Republic
}%


\begin{abstract}
\begin{description}
\item[Abstract] The interaction between freely propagating electrons and light waves is typically described using an approximation in which we assume that the electron velocity remains approximately the same during the interaction. In this article we analytically describe the dynamics of electrons in an interaction potential generated by an optical beat wave beyond this regime and find a structure of sharp electron distribution peaks that periodically alternate in the
energy/momentum spectrum. In the classical description we analytically solve the nonlinear equation of motion, which is an analogy to the mathematical pendulum. While addressing the problem using quantum mechanics, we first use a parabolic approximation of the interaction potential and then we also study the evolution of the electron wavepacket in an infinite periodical potential. Using numerical simulations we show the classical and quantum evolution of the electron spectra during the interaction for different conditions and experimental settings. 
\end{description}
\end{abstract}

\maketitle

\tableofcontents

\section*{Introduction} 
In the early days of quantum mechanics, Kapitza and Dirac came up with theoretical description of a phenomena in which electron waves diffract on an intensity grating formed by a standing electromagnetic wave in vacuum \cite{kapitza1933reflection}. The interaction between electrons and a standing light wave is mediated by the ponderomotive potential which influences the motion of a charged particle in a rapidly oscillating inhomogenous electromagnetic field \cite{chan1979classical, bucksbaum1988high}. In the context of quantum mechanical description, we would be speaking of the stimulated Compton scattering, which is a concept introduced in the original work of Kapitza and Dirac, in which the interacting electron undergoes absorption of a photon from the first wave and subsequently emits a photon into the second wave that together with the first form the mentioned standing wave \cite{efremov1999classical, freimund2001observation, li2004theory, batelaan2007colloquium, dellweg2015kapitza, smirnova2004kapitza}. After the first observations of electron scattering at high-intensity fields in the classical regime \cite{bucksbaum1988high}, the quantum regime of Kapitza-Dirac experiment has been demonstrated both with atoms \cite{Gould1986} and electrons \cite{freimund2001observation}. Since then, the ponderomotive interaction of electrons with light fields has been extensively studied both theoretically \cite{Ahrens2013,Talebi2019,smirnova2004kapitza} and experimentally \cite{Freimund2002,Axelrod2020,kozak2018inelastic,kozak2018ponderomotive,tsarev2023nonlinear} and it has shown a great potential for electron beam shaping \cite{Abajo2021,chirita2022transverse}, electron pulse compression to attosecond durations \cite{hilbert2009temporal,kozak2018ponderomotive,Kozak2019alloptical} or for enhancing the imaging contrast in transmission electron microscopy \cite{Schwartz2019}. 

In this article we study the non-relativistic dynamics of an electron in the ponderomotive potential of an optical beat wave with a focus on the inelastic interaction regime. Inelastic scattering of freely propagating electrons on light waves is
currently a studied topic because of potential applications in advanced electron
microscopy and diffraction \cite{de2010optical, barwick2009photon, baum20094d, park2010photon, ihee2001direct, morimoto2018diffraction, chirita2022transverse}. We rely especially on the generalization of the Kapitza-Dirac effect that describes inelastic scattering of electrons on the ponderomotive potential of a moving optical beat wave \cite{kozak2018inelastic, tsarev2023nonlinear, ebel2024structured, haroutunian1975analogue}. The concept of a temporal lens for electrons was introduced in \cite{hilbert2009temporal} and experimentally used to generate ultrashort electron pulses by utilizing the interaction of electrons with ponderomotive potential of an optical beat wave \cite{kozak2018ponderomotive}. These techniques can be used for modulation and initial characterization of free electron pulses in experiments with subfemtosecond time resolution \cite{hebeisen2006femtosecond, hebeisen2008grating, gao2012full,handali2015creating, kozak2021electron}. It is also worth mentioning that instead of the ponderomotive potential, electrons can also interact with optical near fields of nanostructures \cite{feist2015, piazza2015simultaneous, vanacore2018, kozak2017acceleration, vanacore2019ultrafast, dahan2020resonant,  wang2020coherent, feist2020high, henke2021integrated, shiloh2022quantum, talebi2018electron, talebi2020strong}, which enables us to visualize such fields based on the exchange of momentum between near-field photons and electrons.

In most cases, the inelastic scattering of accelerated electrons on optical fields is described in an approximation of negligible change of the electron's velocity (so-called nonrecoil approximation), however, the effect of recoil in selected processes with slow electrons was also examined \cite{talebi2019near}. It is considered that the acting force depends only on the initial phase and envelope of the field (ponderomotive potential) but does not vary due to the change in position of the electron in the wave's rest frame. This approximation is valid for electrons accelerated to kinetic energies of tens of keV, whose energy changes only by a few eV during the interaction with a femtosecond laser pulse. However, there are applications (e.g. acceleration of electrons using optical fields \cite{breuer2013laser}) that go beyond the applicability of this approximation. In this article we deal with the analytical and numerical solution of electron dynamics in a potential whose shape is given by a harmonic function of the spatial coordinate along the direction of electron propagation or normal to it, beyond the aforementioned approximation.


\section{Classical theory}\label{sec21}
Suppose we have a single electron experiencing harmonic ponderomotive potential (derived in Appendix B)
\begin{equation}\label{eq01:pot}
U_p(z,\,t)=\frac{A}{2}\left(1-\cos\frac{2\pi (v_g t-z)}{\lambda}\right)\,,
\end{equation}
\noindent where $A$ is a constant amplitude, $v_g$ is the potential's group velocity and $\lambda$ is its spatial period. Classical equation of motion will be solved in the potential's rest frame using transformation $q(t)\equiv z(t)-v_g t$. Transformed equation of motion then takes form
\begin{equation}\label{eq02:kyv}
\ddot{q}(t)+\frac{\pi A}{m\lambda}\sin\frac{2\pi q (t)}{\lambda}=0\;.
\end{equation}
\subsection{Solutions to the classical equation of motion}\label{subsec11}
\noindent We recognize the equation \eqref{eq02:kyv} to be a direct analogy of the nonlinear differential equation governing dynamics of mathematical pendulum. We aim to solve the equation in its unsimplified form for general initial conditions
\begin{equation}\label{eq03:initconds}
q(0)=z(0)\equiv z_0\,,\quad\dot{q}(0)=\dot{z}(0)-v_g\equiv\delta v_0\;,
\end{equation}
\noindent whilst formally assuming that relation $|\delta v_0| \ll |v_g|$ holds, which allows us to use \eqref{eq01:pot} as an effective potential experienced even by an electron that is not at rest relative to the potential's rest frame (see Appendix B). We can come by various articles dedicated to finding a solution of the nonlinear equation of motion governing mathematical pendulum, e.g. \cite{belendez2007exact} which considers only the pendulum initially at rest. However, our interest resides in solving the equation for general conditions \eqref{eq03:initconds} because it allows us to include an interaction regime in which the electron's velocity is not synchronized with the group velocity of the ponderomotive potential. Analytical solutions describing electron's classical trajectory will enable us to spot certain interesting aspects, such as the exact period of motion.

The first integration of equation \eqref{eq02:kyv} can be easily carried out, leading towards equation
\begin{equation}
(\dot{q}(t))^2=\frac{A}{m}\cos\frac{2\pi q(t)}{\lambda}+c_1\,.
\end{equation}
\noindent By applying initial conditions \eqref{eq03:initconds}, we can identify the constant of integration $c_1$
\begin{equation}
c_1=\delta v_0^2-\frac{A}{m}\cos\frac{2\pi z_0}{\lambda}\;,
\end{equation}
\noindent so that we may obtain the square of electron's velocity relative to the potential's rest frame in form
\begin{equation}\label{eq04:vkvad}
(\dot{q}(t))^2\equiv\frac{2A}{m}\frac{1}{\kappa^2}\left(1-\kappa^2\sin^2\frac{\pi q (t)}{\lambda}\right)\;,
\end{equation}
\noindent where we have defined a parameter 
\begin{equation}\label{eq05add:kappa}
\kappa^2\equiv \frac{1}{\frac{m\delta v_0^2}{2A}+\sin^2\frac{\pi z_0}{\lambda}}\;.
\end{equation}

Value of the real (and positive) parameter $\kappa$, which is determined by a particular electron's initial conditions $z_0$, $\delta v_0$ and the potential's properties $A$ and $\lambda$, simply indicates whether the electron's trajectory will be bound to a single spatial period of the potential or if its kinetic energy within the potential's rest frame is sufficient for it to travel between periods.

In order to show this, let's assume we have an electron with initial position $z_0$ and velocity in the potential's rest frame $\delta v_0$. The electron's kinetic energy transforms differently from its velocity and we need to take that into account. The kinetic energy relative to the potential's rest frame that is required for the electron to leave the initial spatial period is given by relation
\begin{equation}
\frac{1}{2}m\delta v_0^2>U_p^{max}-U_p(z_0)\,.
\end{equation}
\noindent The potential's maximum $U_p^{max}$ is given by its amplitude $A$. After inserting \mbox{$U_p(z_0)$}, we obtain condition
\begin{equation}
\frac{1}{2}m\delta v_0^2>A-A\sin^2\frac{\pi z_0}{\lambda} \iff 1>\kappa^2\;.
\end{equation}
\noindent We conclude that an electron with $\kappa<1$ will be travelling between spatial periods and an electron with $\kappa>1$ will be trapped within the initial spatial period of the potential. This conclusion will be helpful while solving the equation
\begin{equation}\label{eq06:absfirstder}
|\dot{q}(t)|=\sqrt{\frac{2A}{m\kappa^2}}\sqrt{1-\kappa^2\sin^2\frac{\pi q(t)}{\lambda}}\,.
\end{equation}

We will first solve equation \eqref{eq06:absfirstder} for the case of unbound electrons (in other words $\kappa(z_0,\,\delta v_0,\,\lambda,\,A)<1$), which is arguably simpler. The reason for that is the fact that unbound electron's velocity relative to the potential's rest frame does not change its sign. The sign of $\dot{q}(t)$ is then the same as the sign of $\delta v_0$. We can therefore use separation of variables and after substituting in the integral, we get
\begin{equation}\label{eq07:elliptic}
\int_0^{Q(t)}\frac{d\xi}{\sqrt{1-\kappa^2\sin^2\xi}}=\text{sgn}(\delta v_0)\frac{\pi}{\lambda}\sqrt{\frac{2A}{m\kappa^2}}\:t+c_2\;,
\end{equation}
\noindent where we denoted $Q(t)\equiv\frac{\pi q(t)}{\lambda}$ for compactness. We seek to invert the equation \eqref{eq07:elliptic} in order to get an explicit expression for the trajectory $q(t)$. On the left-hand side of the equation \eqref{eq07:elliptic}, we recognize an incomplete elliptic integral of the first kind $F(Q(t),\,\kappa^2)$. The second constant of integration $c_2$ can be identified from our initial conditions to be
\begin{equation}
c_2=F\left(\frac{\pi z_0}{\lambda},\,\kappa^2\right)\,.
\end{equation}

The task of inverting elliptic integrals is tied to a special class of functions, which are most often called the \textit{Jacobian elliptic functions}. A compact overview of properties of elliptic integrals and their inversions can be found, for example, in \cite{olver2010nist}. One of these functions is the Jacobi amplitude $am (x,\,k)\,$, the definition of which can be formulated as
\begin{equation}\label{eq08:jacobiamp}
x=F(\phi,\,k) \implies \phi= am (x,\,k)\;.
\end{equation}
\noindent Based on \eqref{eq07:elliptic}, we can now express the trajectory of unbound electrons ($\kappa<1$) in our potential's rest frame as
\begin{equation}\label{eq09:unboundtrajec}
q(t)=\frac{\lambda}{\pi} am\left(\text{sgn}(\delta v_0)\frac{\pi}{\lambda}\sqrt{\frac{2A}{m\kappa^2}}\:t+F\left(\frac{\pi z_0}{\lambda},\,\kappa^2\right),\,\kappa^2\right)\;.
\end{equation}

The case of bound electrons ($\kappa>1$) can be solved by converting the problem into one with the electrons ``initially" at rest. The equation \eqref{eq06:absfirstder} for bound electrons then tells us that there will be 2 turning points within the initial spatial period. We can assume $z_0\in\left(-\lambda/2,\,\lambda/2\right)$ and describe these points by
\begin{equation}
\dot{q}(t)=0 \implies |q(t)|=q_{max}\,.
\end{equation}
\noindent The coordinate $q_{max}$ can simply be found as a solution to the equation
\begin{equation}
U_p(q=q_{max})-U_p(q=z_0)=\frac{1}{2}m \delta v_0^2\,.
\end{equation}
\noindent In other words, we are searching for a point in which all the electron's kinetic energy (expressed in the potential's rest frame) was used to rise within the potential. Solving the equation yields the coordinate
\begin{equation}\label{eq010:qmax}
q_{max}=\frac{\lambda}{\pi}\arcsin\frac{1}{\kappa}\;.
\end{equation}
\noindent To find an explicit expression for trajectories of the bound electrons, we will solve the equation \eqref{eq02:kyv} with ``initial" conditions
\begin{equation}\label{eq011:newinit}
q(t=t_0)=q_{max} \implies \dot{q}(t=t_0)=0
\end{equation}
in an unknown time $t_0$. We will subject our discovered solution to the actual initial conditions \eqref{eq03:initconds}, the result of which will be an expression for $t_0$, the time difference between our beginning $t=0$ and the first passage through a turning point $q_{max}$. Equation \eqref{eq04:vkvad} subjected to conditions \eqref{eq011:newinit} then takes form
\begin{equation}\label{eq012:steadypendulum}
(\dot{q}(t))^2=\frac{2A}{m}\left(\sin^2\frac{\pi q_{max}}{\lambda}-\sin^2\frac{\pi q(t)}{\lambda}\right)\,.
\end{equation}
\noindent It is worth to convert this equation, just like in the article \cite{belendez2007exact}, into a form that yields an elliptical integral in \textit{Legendre normal form} after the separation of variables. First of all, we again scale the trajectory $Q(t)\equiv\pi q(t)/\lambda$ for clarity and then we substitute $\phi(t)=\sin Q(t)$ to obtain
\begin{equation}
(\dot{\phi}(t))^2=\frac{2A}{m}\left(\frac{\pi}{\lambda}\right)^2(1-\phi^2(t))(k^2-\phi^2(t))\,,
\end{equation}
\noindent where we denoted $k^2\equiv\sin^2(\pi q_{max}/\lambda)$. An additional scaling $\varphi (t)\equiv\phi(t)/k$ and taking a squareroot of both sides then leads to equation
\begin{equation}\label{eq013:futurelegendre}
\left|\dot{\varphi}(t)\right|=\sqrt{\frac{2A}{m}}\frac{\pi}{\lambda}\sqrt{(1-k^2\varphi^2(t))(1-\varphi^2(t))}\,.
\end{equation}

We demand that the final trajectory satisfies the conditions \eqref{eq03:initconds} and evaluation of \eqref{eq013:futurelegendre} at $t=0$ (after reversing the substitutions) gives us
\begin{eqnarray}
&\frac{\pi}{\lambda k}\cos\frac{\pi z_0}{\lambda}\delta v_0 =&\pm \sqrt{\frac{2A}{m}}\frac{\pi}{\lambda}\sqrt{1-\sin^2\frac{\pi z_0}{\lambda}}\nonumber \\
 & &\times \sqrt{1-\frac{\sin^2\frac{\pi z_0}{\lambda}}{k^2}}\,.
\end{eqnarray}
\noindent Since we know that for $z_0\in (-\lambda/2,\,\lambda/2)$ it holds that $\cos(\pi z_0/\lambda)$ is positive and also
\begin{equation}
k=\sin\frac{\pi q_{max}}{\lambda}>|\sin\frac{\pi z_0}{\lambda}|\,,
\end{equation}
\noindent we conclude that after removing the absolute value, the sign in \eqref{eq013:futurelegendre} will once again be given by  $\text{sgn}(\delta v_0)$. After separation of variables and integration of both sides, the equation then takes form
\begin{equation}
\int_{\varphi(t_0)}^{\varphi(t)}\frac{d\theta}{\sqrt{(1-k^2\theta^2)(1-\theta^2)}}=\text{sgn}(\delta v_0)\sqrt{\frac{2A}{m}}\frac{\pi}{\lambda}(t-t_0)\,,
\end{equation}
\noindent where we can realize from the conditions \eqref{eq011:newinit} that
\begin{equation}
\varphi(t_0)=\frac{\sin\frac{\pi q_{max}}{\lambda}}{k}=1\,.
\end{equation}
\noindent Splitting the boundaries of integration and identification of the corresponding elliptical integrals (we must take into account that these elliptical integrals are in Legendre normal form) leads to equation
\begin{equation}
F(\arcsin\varphi(t),\,k^2)=K(k^2)+\text{sgn}(\delta v_0)\sqrt{\frac{2A}{m}}\frac{\pi}{\lambda}(t-t_0)\,,
\end{equation}
\noindent where $K(k^2)$ is the complete elliptical integral of the first kind with parameter $k^2$. If we evaluate the equation at $t=0$ and insert the conditions \eqref{eq03:initconds}, we get the expression for $t_0$ in form
\begin{eqnarray}
&t_0=&\text{sgn}(\delta v_0) \frac{\lambda}{\pi}\sqrt{\frac{m}{2A}}\nonumber \\
 & &\times\left(K(k^2)-F\left(\arcsin\left(\frac{\sin\frac{\pi z_0}{\lambda}}{k}\right),\,k^2\right)\right)\,.
\end{eqnarray}
\noindent The remaining task is then to invert relation
\begin{eqnarray}\label{eq014:lastbeforeinverse}
&F(\arcsin\varphi(t),\,k^2)&=\text{sgn}(\delta v_0)\sqrt{\frac{2A}{m}}\frac{\pi}{\lambda}\,t\nonumber \\
 & &+ F\left(\arcsin\left(\frac{\sin\frac{\pi z_0}{\lambda}}{k}\right),\,k^2\right)\,,
\end{eqnarray}
\noindent for which we will use another one of the Jacobi elliptic functions, namely $sn(x,\,m)$ (can be found in \cite{olver2010nist}). This function can be defined in two following (equivalent) ways 
\begin{align}
F(\arcsin\phi,\,m)=x \implies \sin\phi=sn(x,\,m)\;,\\sn(x,\,m)=\sin am(x,\,m)\;.
\end{align}

\noindent In the interest of compactness, we should realize that the parameters $k$ and $\kappa$ are not independent. In fact, it is evident from \eqref{eq010:qmax} that they are related in a simple way
\begin{equation}
k=\sin\frac{\pi q_{max}}{\lambda}=\frac{1}{\kappa}\;.
\end{equation}
\noindent Inverting the equation \eqref{eq014:lastbeforeinverse} then leads, after reversing our substitutions, to the analytical expression for trajectories of bound electrons ($\kappa>1$) in the rest frame of our potential
\begin{eqnarray}\label{eq015:boundsolution}
&q(t)=&\frac{\lambda}{\pi}\arcsin\left\{\frac{1}{\kappa}sn\left[\text{sgn}(\delta v_0)\frac{\pi}{\lambda}\sqrt{\frac{2A}{m}}\:t\right.\right.\nonumber \\
 & &+\left.\left.F\left(\arcsin\left(\kappa\sin\frac{\pi z_0}{\lambda}\right),\,\frac{1}{\kappa^2}\right),\,\frac{1}{\kappa^2}\right]\right\}\,.
\end{eqnarray}

\subsection{Selected properties of the classical trajectories}\label{subsec12}
Acquired analytical solutions \eqref{eq09:unboundtrajec} and \eqref{eq015:boundsolution} allow us to perform precise simulations, in order to get the expected dynamics of the electron distribution. It is therefore neccessary to select suitable values of physical quantities. That is precisely why we devote a few paragraphs to certain aspects of the classical electron's dynamics, such as the maximal change of electron's kinetic energy and the appropriate interpretation of the condition $\kappa>1$ for binding electrons within one spatial period of the potential.

Classical approximation of the studied interaction is appropriate if the longitudinal coherence length of the electron beam is significantly shorter than the potential's spatial period \cite{pan2023weak}. We may therefore assume for simplicity that the initial spatial distribution of electrons is constant within the interval $\langle -\frac{\lambda}{2},\,\frac{\lambda}{2}\rangle$ in order to describe all possible initial configurations. Initial electron distribution in the momentum space will be characterized by the offset of kinetic energy (expressed in the laboratory frame) compared to the synchronous case, where $\delta v_0=0$. Mentioned initial offset is then

\begin{eqnarray}\label{eq016:delE0}
&\Delta E_0& \equiv \frac{1}{2}\,m\left[(v_g+\delta v_0)^2 - v_g^2\right]\nonumber \\
 & &=\frac{m\delta v_0}{2}(\delta v_0 + 2v_g)\,.
\end{eqnarray}
\noindent Solving for $\delta v_0$ yields
\begin{equation}
\delta v_0=-v_g\pm\sqrt{v_g^2+\frac{2\Delta E_0}{m}}\,.
\end{equation}
\noindent Since we assume $|\delta v_0|\ll |v_g|$, we know that only the expression with sign $+$ is applicable.

The periodicity and classification (bound vs. unbound) of discovered trajectories is discussed in the Appendix A. Therefore, we know that regardless of whether the electron is bound or unbound, its velocity relative to the potential's rest frame will oscillate (pseudoperodical unbound trajectory shifts by a constant, its derivative is thus periodical). The time-dependent offset of kinetic energy
\begin{equation}\label{eq017:deltaE(t)}
\Delta E(t) = \frac{1}{2}m \left[(v_g+\dot{q}(t))^2-v_g^2\right]\,,
\end{equation}
\noindent will then likewise oscillate with a period which we can determine from our analytical solutions. The boundary values of $\Delta E(t)$, or the minimal and maximal kinetic energy of an electron during the interaction, can be simply determined from the first integral of the equation of motion in the form \eqref{eq06:absfirstder}. We know that bound electrons ($\kappa >1$) have 2 turning points within the potential. Contrary to that, the unbound electrons ($\kappa < 1$) are never at rest relative to the potential's rest frame. Based on the equation \eqref{eq06:absfirstder}, we can therefore summarize the boundary values $\dot{q}_{min}$ and $\dot{q}_{max}$, seen in table \ref{tab1:velocity}.

\begin{table}[h!]
\catcode`\-=12
\caption{\justifying Table summarizing the boundary values of velocity relative to the potential's rest frame for bound ($\kappa > 1$) and unbound ($\kappa < 1$) electrons, which are initially faster ($\delta v_0>0$) or slower ($\delta v_0<0$) than the potential.}
\centering
\begin{tabularx}{\columnwidth}{@{}c*{4}{>{\centering\arraybackslash}X}@{}}
\toprule
                & \multicolumn{2}{c}{$\kappa>1$}                                            & \multicolumn{2}{c}{$\kappa<1$}                                                        \\ \midrule
                & $\dot{q}_{min}$              & $\dot{q}_{max}$             & $\dot{q}_{min}$                   & $\dot{q}_{max}$                    \\ \cmidrule(l){2-5} 
$\delta v_0 >0$ & $-\frac{1}{\kappa}\sqrt{\frac{2A}{m}}$ & $\frac{1}{\kappa}\sqrt{\frac{2A}{m}}$ & $\sqrt{\frac{2A}{m}\frac{1-\kappa^2}{\kappa^2}}$ & $\frac{1}{\kappa}\sqrt{\frac{2A}{m}}$        \\ \cmidrule(r){1-1}
$\delta v_0 <0$ & $-\frac{1}{\kappa}\sqrt{\frac{2A}{m}}$ & $\frac{1}{\kappa}\sqrt{\frac{2A}{m}}$ & $-\frac{1}{\kappa}\sqrt{\frac{2A}{m}}$      & $-\sqrt{\frac{2A}{m}\frac{1-\kappa^2}{\kappa^2}}$ \\ \bottomrule
\end{tabularx}

\label{tab1:velocity}
\end{table}

We can see for example that if we take a bound electron, which has been initially slower than the potential's rest frame, the maximum value of its offset of kinetic energy will be given by \footnote{One must not forget that $\kappa\equiv\kappa(z_0,\,\Delta E_0)\,.$ }
\begin{equation}
\Delta E_{max} = \frac{A+\kappa v_g \sqrt{2Am}}{\kappa^2}\,.
\end{equation}

If we take a look at the function $\kappa\equiv\kappa(z_0,\,\Delta E_0)$ (seen at \eqref{eq05add:kappa}), we can conclude that for arbitrary initial conditions $z_0,\,\delta v_0$ (except \mbox{$z_0=\pm\lambda/2$)} there is a critical value of the amplitude $A_{crit}$ that allows us to achieve binding of the electron. The critical value $A_{crit}$ can be expressed by solving the condition $\kappa=1$ for the amplitude (after putting in $\delta v_0\, (\Delta E_0)$ ), which yields 
\begin{eqnarray}\label{eq018:Akrit}
&A_{crit} (z_0, \Delta E_0)&=\frac{m}{2\cos^2\left(\frac{\pi z_0}{\lambda}\right)}\nonumber \\
 & &\times\left(\sqrt{v_g^2+\frac{2\Delta E_0}{m}}-v_g\right)^2\;.
\end{eqnarray}
\noindent This relation allows us to choose an appropriate amplitude $A$ for a set of initial conditions $z_0,\,\Delta E_0$ (with an arbitrary distribution), while taking into account the corresponding expected fraction of electrons bound within one spatial period of the potential.

Based on a considered set of initial conditions and the selected amplitude $A$, we can also construct a normalized distribution function (probability density) for the period $T$ of the energy ($\Delta E (t)$) oscillations. $T$ can be characterized in terms of the complete elliptical integral of the first kind. For a given potential, we can calculate the period as \footnote{A direct consequence of the analysis seen in \hyperref[append]{Appendix A}.}
\begin{equation}\label{eq019:period}
T=\begin{cases} \;\frac{\lambda \kappa}{\pi}\sqrt{\frac{2m}{A}}\,K(\kappa^2)\,,\quad \text{if}\; \kappa(z_0,\,\Delta E_0)<1\,, \\
\;\frac{2\lambda}{\pi}\sqrt{\frac{2m}{A}}\,K\left(\frac{1}{\kappa^2}\right),\quad \text{if}\; \kappa(z_0,\,\Delta E_0)>1\,. \end{cases}
\end{equation}
\noindent We will later take a set of pairs of initial conditions, $z_0$ with a constant distribution and $\Delta E_0$ from a narrow gaussian distribution, we will see that there are two significant peaks in the probability density for the period $T$. The peak over smaller $T$ will correspond to a significant portion of the unbound electrons and the one over larger $T$ will belong to the bound electrons. Their relative size and sharpness are always dictated by the fraction of bound electrons (in other words, by the considered amplitude $A$). It will also allow us to recognize that the most probable period for bound electrons is only slightly shifted from the period of linearized mathematical pendulum (in our case $T_{lin}=\lambda \sqrt{2m/A}$). 
\section{Quantum treatment}
In this chapter, we will deal with the interaction of an electron with the ponderomotive potential using quantum mechanical description. As for the potential iself, we will assume its classical form \eqref{eq01:pot} and describe the evolution only in its rest frame. Considered Hamiltonian of our system therefore takes the form
\begin{equation}\label{eq021:operators}
\hat H = \frac{\hat{p}^2}{2m}+\frac{A}{2}\left(\hat{1}-\cos\frac{2\pi \hat{q}}{\lambda}\right)\,,
\end{equation} 
The first approximation will adress a situation in which the initial wavefunction is localized around the potential's minimum and the dynamics is only insignificantly influenced by the other spatial periods of the potential. Therefore, it will be handy to use a parabolic approximation for the potential. The problem of a full cosine potential is going to be studied for an electron wavefunction in the form of a plane wave, using the Bloch's theorem and a suitable unitary transformation.

\subsection{Parabolic approximation}
At first, we take the mentioned situation in which the initial electron wavefunction is localized around the potential's minimum and it stays this way for the duration of the interaction. \footnote{It might seem slightly unneccessary, since it means that the coherence length of initial electron beam is smaller than the potential's spatial period, therefore enabling classical description. Indeed, we will be able to see that the final results bear a resemblance to the classical case. } These assumptions limit us to an interaction regime in which we are able to reliably utilize the parabolic approximation
\begin{equation}
V(\hat{q})=\frac{A}{2}\left(\hat{1}-\cos\frac{2\pi \hat{q}}{\lambda}\right)\approx \frac{A\pi^2}{\lambda^2}\hat{q}^2\,.
\end{equation}
\noindent If we denote
\begin{equation}
\Omega=\frac{\pi}{\lambda}\sqrt{\frac{2A}{m}}\,,
\end{equation}
\noindent we get the well-known Hamiltonian of linear harmonic oscillator
\begin{equation}\label{eq023:quantumoscilator}
\hat{H}=\frac{\hat{p}^2}{2m}+\frac{1}{2}m\Omega^2\hat{q}^2\,.
\end{equation}

Solution to the stationary Schrödinger equation can be, in case of Hamiltonian \eqref{eq023:quantumoscilator}, found in every introductory book of quantum mechanics. We find discreet eigenvalues $E_n$, where $n\in \mathbb{N}\cup\{0\}$, with corresponding wavefunctions $\psi_n (x)$ of the state eigenvectors in form
\begin{align}\label{eq024:LHOsol}
E_n=& \hbar \Omega\left(n+\frac{1}{2}\right)\,,\\
\psi_n (q)=& \frac{\sqrt[4]{\frac{m\Omega}{\pi \hbar}}}{\sqrt{2^n\, n!}}e^{-\frac{m\Omega}{2\hbar}q^2}H_n\left(q\sqrt{\frac{m\Omega}{\hbar}}\right)\;.
\end{align}
\noindent $H_n (x)$ denotes the \textit{Hermite polynomial} in variable $x$, which can be expressed as
\begin{equation}
H_n(x)=(-1)^n e^{x^2}\frac{d^n}{dx^n}(e^{-x^2})\,.
\end{equation}
\noindent The set of wavefunctions $\psi_n (q)$ constitutes a complete orthonormal basis of the Hilbert space $\mathcal{H}$, which is realized by $L^2(\mathbb{R})$ for the physical system of linear harmonic oscillator.

From the knowledge that initial state $|\phi_0\rangle\equiv|\psi(0)\rangle$ evolves as
\begin{equation}
|\psi(t)\rangle=e^{-\frac{i}{\hbar}\hat{H}t}|\phi_0\rangle\,,
\end{equation}
\noindent we can therefore write the time-dependent wavefunction in the form of a series
\begin{align}\label{eq027:dynamicssol}
\psi(t,\,q)=\sum_{n=0}^\infty c_n e^{-\frac{i}{\hbar}E_n t}\,\psi_n(q)\,,\\
c_n=\int_{-\infty}^\infty \psi_n^\ast(q)\,\phi_0(q)\,dq\,.
\end{align}
\subsection{Plane wave in full harmonical potential}\label{sec22}
Let's address the physical situation in which the longitudinal coherence length of our electron beam is comparable or larger than the spatial period of considered ponderomotive potential. In this interaction regime, we can no longer utilize the parabolic approximation since the electron experiences a full periodical potential. It is also expected that precisely this regime will display a significant quantum nature of the interaction. We will, once again, describe the dynamics in the potential's rest frame, but this time we will consider Hamiltonian in the form \footnote{The absence of a term that would be linear in the field's vector potential is caused by the fact that the net contribution of an oscillating vector potential to the static field vanishes after integration.}
\begin{equation}\label{eq028:periodicham}
\hat{H}=\frac{\hat{p}^2}{2m}+\frac{A}{2}\left(\hat{1}-\cos\frac{2\pi\hat{q}}{\lambda}\right)\,.
\end{equation}

In order to solve the Schrödinger equation, we will recall Bloch's theorem, which states that for a general periodical potential $V(\bm r + \bm R)=V(\bm r)$, where $\bm R$ is the \textit{lattice vector}, one can assume that the sought eigenfunctions are in the form
\begin{equation}\label{eq029:blochfunc}
\psi_{n\bm k}(\bm r)=e^{i\bm k \boldsymbol{\cdot} \bm r}\,u_{n\bm k}(\bm r)\,.
\end{equation}
\noindent Functions $u_{n\bm k}(\bm r)$ are periodical with respect to the lattice vector $\bm R$, in other words $u_{n\bm k}(\bm r + \bm R)=u_{n\bm k}(\bm r)$. Energy spectrum of an idealized infinitely periodical potential is neccessarily continuous and has a band structure (see e.g. \cite{cejnar2013condensed}). The eigenfunctions $\psi_{n\bm k}(\bm r)$ are not normalizable (in the standard sense), since they are not integrable. Vector $\bm k$, usually called the \textit{crystal momentum} or \textit{quasimomentum}, represents a quantum number that describes a specific state within a given band. Vector $\bm k$ does not describe a unique state, for that we must specify a value of the discrete index $n$, which is most commonly referred to as the \textit{band index}.

A suitable way of solving the eigenproblem is to take advantage of the symmetry of quantum description under a transformation, which can be connected to a unitary operator $\hat{U}$. If we choose $\hat{U}=e^{-i\bm k \boldsymbol{\cdot}\hat{\bm r}}$ , we acquire \footnote{The equality $e^{-i\bm k \boldsymbol{\cdot}\hat{\bm r}} \hat{\bm p}\, e^{+i\bm k \boldsymbol{\cdot}\hat{\bm r}}=\hat{\bm p}+\hbar \bm k$ is a direct consequence of the \textit{BCH formula}  and cannonical commutation relations $\left[\hat{\bm r}_m,\, \hat{\bm p}_n\right] = i\hbar \delta_{mn}$\,. }
\begin{align}\label{eq030:gaugetransform}
\psi'_{n\bm k}(\bm r)&=u_{n\bm k}(\bm r)\,,\\
\hat{H}_{\bm k}'&=\frac{\left(\hat{\bm p}+\hbar \bm k\right)^2 }{2m} + V(\hat{\bm r})\,.
\end{align}

We wish to apply Bloch's theorem and the mentioned transformation in order to solve the stationary one-dimensional Schrödinger equation
\begin{equation}
\left[-\frac{\hbar^2}{2m}\frac{d^2}{dq^2}+\frac{A}{2}\left(1-\cos\frac{2\pi q}{\lambda}\right)\right]\psi (q) = E \psi (q)\,.
\end{equation}
\noindent However, our main goal is to acquire predictions for the dynamics of an electron with a specific initial state $|\phi_0\rangle$. We would therefore like to decompose the initial state into the eigenstates of our Hamiltonian. Bloch ansatz for the eigenfunctions \eqref{eq029:blochfunc} with a general $\bm k$ (in 1D only one component $k$) would be suitable if we wanted to study the band structure (dispersion relations $E_n(k)$), but it would be rather difficult to acquire suitable approximations for eigenfunctions, even without considering that our intention is to solve the non-stationary problem.

We will consider the initial electron wavefunction in the form of a plane wave $\phi_0 (q) =\frac{1}{\sqrt{\lambda}} e^{ik_0q}$, in other words, we will examine the evolution of a sharp momentum state. In order to pave the way for numerical calculations, we will choose \footnote{A step which is subsequently justified by the convergence rate of decomposition coefficients.} $k\equiv k_0$ in the ansatz  \eqref{eq029:blochfunc} which limits the eigenproblem to a discrete case
\begin{eqnarray}\label{eq031:botchedbloch}
\left[-\frac{\hbar^2}{2m}\frac{d^2}{dq^2}+\frac{A}{2}\left(1-\cos\frac{2\pi q}{\lambda}\right)\right]\psi_n (q)\nonumber\\ = E_n \psi_n (q)\,,
\end{eqnarray}
\noindent where $\psi_n (q) = e^{ik_0 q}\,u_n(q)$ and $u_n(q+\lambda)=u_n(q)\,$. If we now perform a unitary transformation, analagous to the general case that led us to \eqref{eq030:gaugetransform}, we acquire the equation
\begin{eqnarray}\label{eq032:botchedgauge}
\left[-\frac{\hbar^2}{2m}\left(\frac{d}{dq}+ik_0\right)^2+\frac{A}{2}\left(1-\cos\frac{2\pi q}{\lambda}\right)\right]u_n(q)\nonumber\\=E_n u_n(q)\,.
\end{eqnarray}
\noindent The most significant advantage of transforming the problem in this way is that now we are seeking only the $\lambda$-periodical functions $u_n(q)$, therefore we are able to limit ourselves to the space of quadraticaly integrable functions $L^2(\langle0,\,\lambda\rangle)$ with the complete orthonormal basis $\phi_j(q)=\frac{1}{\sqrt{\lambda}}e^{\frac{2\pi i j}{\lambda}q}\,,\,j\in \mathbb{Z}\,$. This allows us to assume the decomposition
\begin{align}
&u_n(q)=\sum_{j=-\infty}^\infty c_{nj}\phi_j(q)\,,
\\
&c_{nj}\equiv\int_0^\lambda\frac{1}{\sqrt{\lambda}}e^{-\frac{2\pi i j}{\lambda}q}\:u_n(q)\,dq\,.   
\end{align}
\noindent Inserting this decomposition into the equation \eqref{eq032:botchedgauge}, multiplying by $\phi^\ast_l(q)$ and integrating both sides over the interval $\langle 0,\,\lambda\rangle$ leads (due to orthonormality of $\phi_j(q)$) to the equation
\begin{equation}\label{eq033:infinitematrixprob}
\sum_{j=-\infty}^\infty H'_{lj}c_{nj}=E_n c_{nl}\,.
\end{equation}
$H'_{lj}$ denotes a matrix element of the transformed Hamiltonian in the basis $\{\phi_j(q)\}_{j=-\infty}^\infty\:$, which can be evaluated analytically. The result is a tridiagonal matrix with elements
\begin{eqnarray}\label{eq034:matrixelem}
&H'_{lj}=&\frac{1}{2}\left(A+\frac{\hbar^2}{m\lambda^2}(2\pi j+k_0\lambda)^2\right)\delta_{lj}\nonumber \\
 & &-\frac{A}{4}(\delta_{l,\,j+1}+\delta_{l,\,j-1})\,.
\end{eqnarray}

We have therefore acquired a clear method for solving the equation numerically. It is self-evident that it would not be simple to find an exact expression for $u_n(q)$, since it would require the construction and diagonalization of an infinitely-dimensional matrix $H'_{lj}$ with diverging diagonal elements. However, for now we assume to have the decomposition coefficients $c_{nj}$ and eigenvalues for energy $E_n\,$ at our disposal. Considered initial electron wavefunction $\phi_0(q)=\frac{1}{\sqrt{\lambda}}e^{ik_0q}$ has a simple form of a constant in the transformed reference frame $\phi_0'(q)=\frac{1}{\sqrt{\lambda}}$. Its decomposition into the eigenfunctions $u_n(q)$ therefore yields

\begin{eqnarray}
&\phi_0'(q)& = \sum_{n=0}^\infty \kappa_n u_n(q)
\\
&\kappa_n& \equiv\int_0^\lambda \frac{1}{\sqrt{\lambda}} u_n^\ast(q)\,dq  =c_{n0}^\ast\,.
\end{eqnarray}
\noindent According to relation \eqref{eq027:dynamicssol}, where $u_n (q)$ now play the role of eigenfunctions corresponding to energies $E_n$, the resulting time-dependent wavefunction can be written in the transformed reference frame as
\begin{eqnarray}
&\psi' (t,\,q)&=\sum_{n=0}^\infty c_{n0}^\ast\, e^{-\frac{i}{\hbar}E_n t}\,u_n(q)\nonumber \\
 & & = \sum_{n=0}^\infty c_{n0}^\ast\, e^{-\frac{i}{\hbar}E_n t}\sum_{j=-\infty}^\infty \frac{c_{nj}}{\sqrt{\lambda}} e^{\frac{2\pi i j}{\lambda}\,q}\,.
\end{eqnarray}
\noindent Solution in the original frame, which is nothing else than the potential's rest frame, can therefore be expressed in form
\begin{eqnarray}\label{eq035:BlochSol}
\psi(t,\,q)&=&\sum_{n=0}^\infty \sum_{j=-\infty}^\infty \frac{c_{n0}^\ast c_{nj}}{\sqrt{\lambda}} \nonumber \\
 & &\times\exp\left(ik_0 q +\frac{2\pi i j}{\lambda}\,q-\frac{i}{\hbar}\,E_n t \right)\,.
\end{eqnarray}
\noindent We can also acquire the momentum representation of this wavefunction as
\begin{eqnarray}\label{eq036:BlochSolMomentum}
\Psi(t,\,p)&=&\sum_{n=0}^\infty \sum_{j=-\infty}^\infty \sqrt{\frac{2\pi\hbar}{\lambda}}\,c_{n0}^\ast c_{nj}\nonumber \\
 & &\times e^{-\frac{i}{\hbar}E_n t}\,\delta\left[p-\hbar\left(k_0+\frac{2\pi j}{\lambda}\right)\right]\,.
\end{eqnarray}

Expression \eqref{eq036:BlochSolMomentum} for the resulting wavefunction shows that for an initial state \mbox{$\Phi_0(p)\propto\delta(p-\hbar k_0)$} we have acquired a superposition of discrete momentum eigenstates, which are seperated by an integer multiple of $\Delta p = 2\pi \hbar /\lambda$. For the assumed case of ponderomotive potential being generated by two counterpropagating plane waves of different frequencies $\omega_1$ and $\omega_2$ (seen in Appendix B), we know that \mbox{$\lambda=2\pi c /(\omega_1+\omega_2)$,} therefore $\Delta p = \hbar (\omega_1+\omega_2)/c$. We then see that the spacing of discrete momentum levels corresponds to the electron's absorption of a photon from one of the optical plane waves and emitting a photon to the other plane wave due to the \textit{stimulated Compton scattering}.  Based on this observation, let us introduce the time-dependent occupancy rate of the momentum level with index $j$ as
\begin{equation}
\Sigma_j (t)\equiv\sum_{n=0}^\infty c^\ast_{n0} c_{nj} e^{-\frac{i}{\hbar}E_n t}\,,
\end{equation}

\noindent which will ultimately be the quantity of interest when discussing numerical simulations.

\section{Results of numerical simulations}
This chapter will cover a simple numerical analysis of acquired analytical results, as well as the visualization of some predicted aspects of the dynamics of electron distributions using specific sets of physical parameters. More attention will be given to a thorough discussion of classical results as they offer a keener insight into the behaviour and physical limitations of the interaction.
\subsection{Dynamics based on classical trajectories} \label{chapter3.1}
In order to visualize the evolution of a specific electron distribution, we need to find the appropriate values of physical parameters. If we are given a set of initial conditions for electron trajectories, we can use relation \eqref{eq018:Akrit} to numerically determine dependance of the fraction of bound electrons on amplitude $A$. 

Regarding the choice of spatial scale, suppose that our harmonic ponderomotive potential is in fact created by two counterpropagating monochromatic waves of different frequencies, which could be generated as a second and third harmonic frequency of a base femtosecond laser pulse with wavelength $\lambda_0$. Then it holds that the spatial period of our ponderomotive potential will be $\lambda=\lambda_0/5$ and its group velocity will be $v_g=c/5$. In all of the simulations, we will asume that the original femtosecond laser pulse has the wavelength $\lambda_0=1030\rm\: nm$, thus fixing the potential's spatial period \footnote{Since the amplitude in this setup still depends on intensity of the original laser pulse, we have fixed only the spatial period.} to $\lambda = 206\:\rm nm$.

For simplicity it will be assumed that the initial electron positions are sampled from a constant distribution over one spatial period of the potential and the initial offsets of kinetic energy (defined as \eqref{eq016:delE0}) are sampled from a Gaussian distribution, characterized by its mean value $\mu_E$ and its full width at half maximum ${\Delta E}_{\text{FWHM}}$. Our goal is to demonstrate the qualitative difference between the synchronous and asynchronous case, therefore the notation $\mu_E^{(s)}=0$ and $\mu_E^{(a)}\neq0$ will be used. We will also see that for higher values of $|\mu_E^{(a)}|$ a significantly larger $A$ is required in order to achieve the same fraction of bound electrons. Because of that, we will also distinguish between weak and strong interaction regime. However, this classification is primarily in the interest of intuition when it comes to maximal kinetic energy offsets and oscillation periods. For equal bound electron fractions with the same shapes of initial condition distributions, the spectra are qualitatively the same.

\begin{figure}[b!]\centering
\includegraphics[scale=0.85]{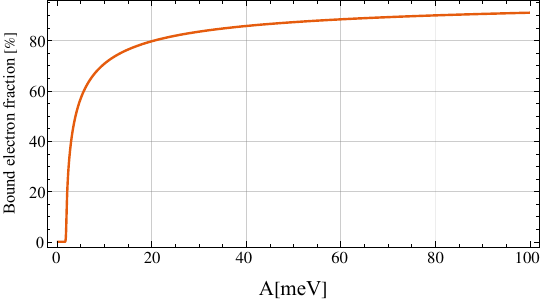}
\caption{\justifying Dependance of bound electron fraction on amplitude of the potential for Gaussian distribution of $\Delta E_0$ with $\mu_E^{(a)}=-9\,\rm eV$ and ${\Delta E}_{\text{FWHM}}=0.5\,\rm eV$.}
\label{obr01:weakcapturedelectronfraction}

\end{figure}

\begin{figure}[b!]
\centering
\begin{subfigure}{0.35\textwidth}
  \centering
  \includegraphics[width=\textwidth]{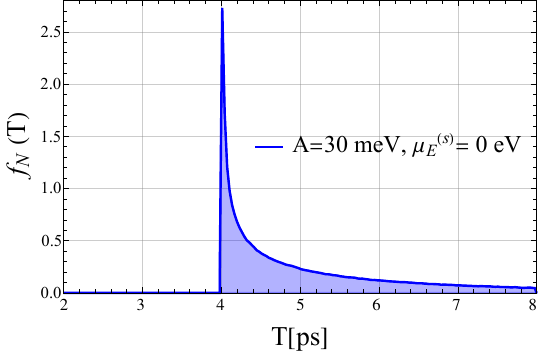}
  \caption{}
  \label{fig:sub1}
\end{subfigure}%
\hfill
\begin{subfigure}{0.35\textwidth}
  \centering
  \includegraphics[width=\textwidth]{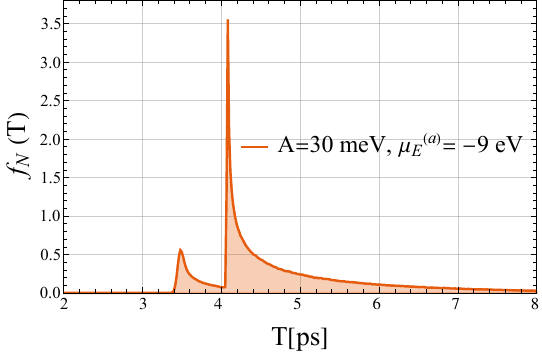}
  \caption{}
  \label{fig:sub2}
\end{subfigure}
\caption{\justifying Normalized distribution functions $f_N (T)$ of period $T$ on the electron ensemble for \textbf{(a)} synchronous case $\mu_E^{(s)}=0\,\rm eV$ and \textbf{(b)} asynchronous case  $\mu_E^{(a)}=-9\,\rm eV$ with identical widths of energy distributions ${\Delta E}_{\text{FWHM}}=0.5\,\rm eV$ and amplitudes $A=30\,\rm meV$.}
\label{obr02:periods}
\end{figure}

To address the weak interaction, we will choose (in accordance with \cite{kozakarticle}) mean value for asynchronous energy distribution $\mu_E^{(a)}=-9\,\rm eV$ with width ${\Delta E}_{\text{FWHM}}=0.5\,\rm eV$ that will also be assumed for the synchronous case so we could compare their energy spectra evolutions. In order to find an appropriate amplitude for our potential, we first plot the dependance of bound electron fraction on $A$. This can easily be done by calculating the relative cumulative frequency of $A_{crit}$ over the set of initial conditions $\{(z_0,\,\Delta E_0)_i)\}_{i=1}^N$, where $N$ is the total number of electrons (plot seen in figure \ref{obr01:weakcapturedelectronfraction}). Figure \ref{obr01:weakcapturedelectronfraction} allows us to observe the quick rise and subsequent saturation of bound electron fraction with increasing potential's amplitude, we will therefore choose $A= 30 \,\rm meV$ which corresponds \mbox{to $\approx 85\:\%$} of bound electrons.

\begin{figure}[b!]
\centering
\begin{subfigure}{0.48\textwidth}
  \centering
  \includegraphics[width=\textwidth]{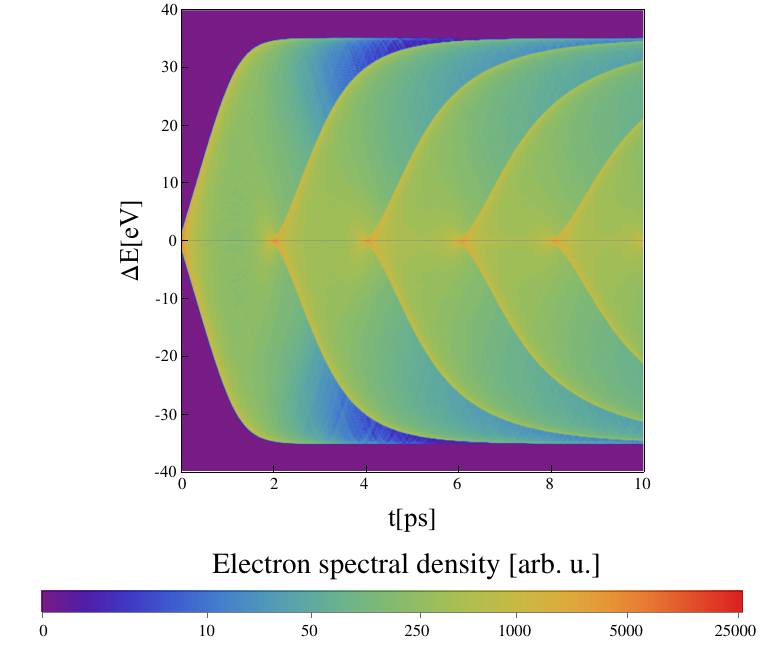}
  \caption{}
  \label{obr03a:energyspectrumsynch}
\end{subfigure}%
\hfill
\begin{subfigure}{0.48\textwidth}
  \centering
  \includegraphics[width=\textwidth]{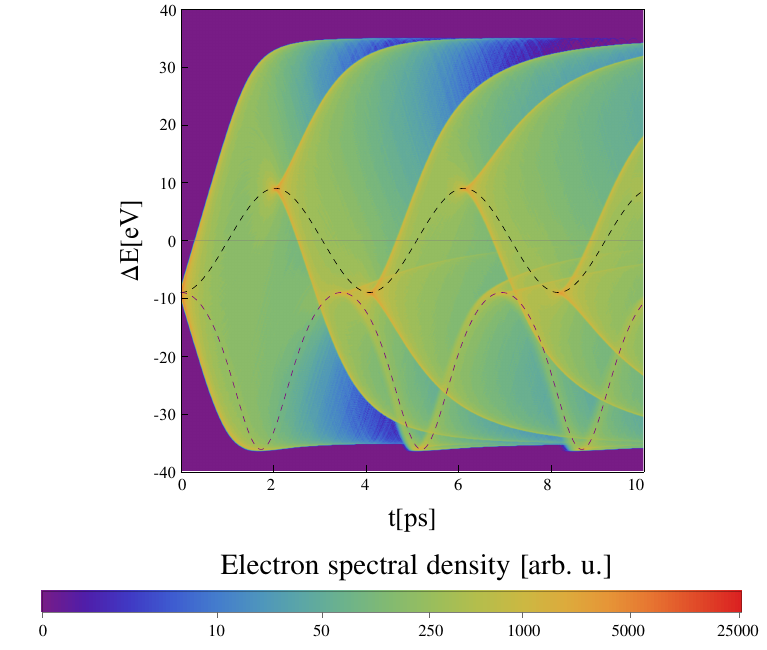}
  \caption{}
  \label{obr03b:energyspectrumasynch}
\end{subfigure}
\caption{\justifying Evolution of electron spectral densities in energy space parametrized by the offset of kinetic energy $\Delta E$ with initial Gaussian distribution width \mbox{${\Delta E}_{\text{FWHM}}=0.5\,\rm eV$} and potential amplitude $A=30\, \rm meV$ in \textbf{(a)} synchronous case \textbf{(b)} asynchronous case with mean value $\mu_E^{(a)}=-9\,\rm eV$. Black dashed line represents a typical bound electron trajectory ($z_0=0$) and purple dashed line a typical unbound electron trajectory ($z_0=\lambda/2$).}
\label{obr03:energyspectra}
\end{figure}

We can use the set of initial conditions for electron trajectories and assumed value of $\lambda$ and $A$ to plot a probability density (or normalized distribution function) $f_N (T)$ for period $T$ of oscillations in energy space, which will be useful for interpreting the dynamics of electron spectra. Plotted probability densities for synchronous and asynchronous case (seen in figure \ref{obr02:periods}) show one main peak for synchronous interaction, which is exactly period $T_{lin}=\lambda \sqrt{2m/A}\,$ of linearized mathematical pendulum. Asynchronous case shows a slightly shifted main peak (due to non-linearity of the trajectories) and also the emergence of a smaller, less sharp peak over smaller $T$, which corresponds to a large portion of the unbound electrons. 

\begin{figure}[t!]\centering
\includegraphics[scale=0.47]{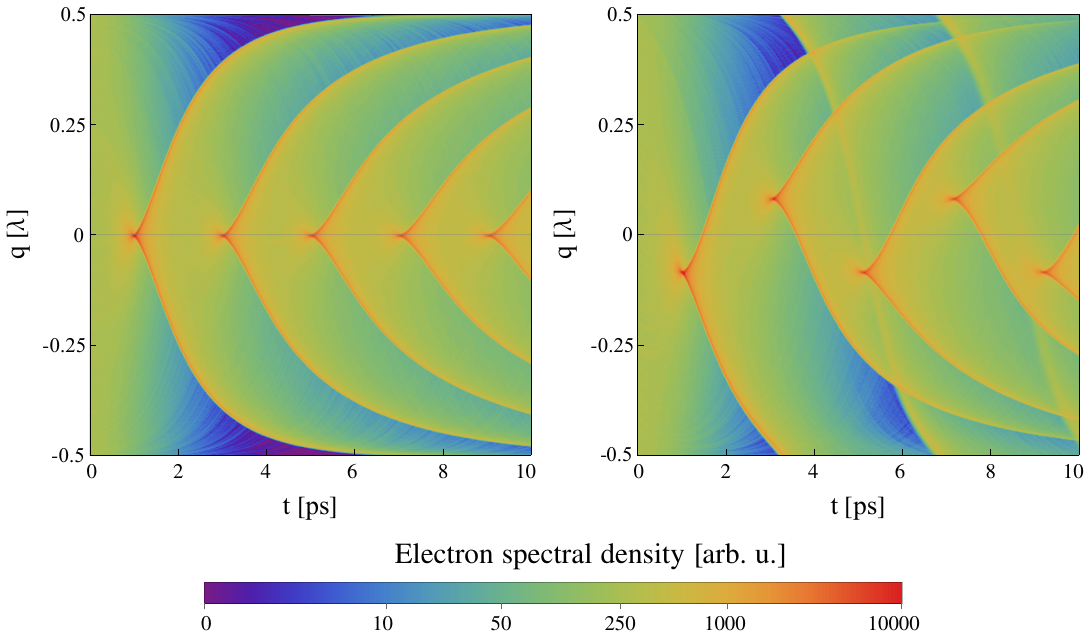}
\caption{\justifying Evolutions of electron spectral densities in real space in the synchronous (left) and asynchronous (right) regime with parameters corresponding to the situation visualized on figures \ref{obr03a:energyspectrumsynch} and \ref{obr03b:energyspectrumasynch} respectively.}
\label{obr04:coordspectra}

\end{figure}

Figure \ref{obr03:energyspectra} is a result of numerical simulation of $10^5$ electrons, for each of which we calculate $\kappa(z_0,\,\Delta E_0)$ and then use the corresponding expression for trajectory, the derivative of which can be expressed either as another Jacobi elliptic function or directly obtained from \eqref{eq06:absfirstder}. Both the synchronous and asynchronous interaction regime contain periodical \footnote{Apparent half period in the synchronous case is naturally caused by symmetry of the trajectory analogous to a mathematical pendulum released without initial velocity.} focusing and blurring of the electron spectral density \footnote{Comparison with figure \ref{obr04:coordspectra} allows us to recognize that foci of the distribution in energy space correspond to its defocusing in the real space and vice versa.}. However, a more interesting dynamics happens in the asynchronous regime \ref{obr03b:energyspectrumasynch} where initial blurring of the spectrum is followed by periodical emergence of foci alternating between positive and negative values of $\Delta E$. Logarithmic scaling of contours also allows us to observe a fraction of electron distribution oscillating faster only at the negative part of $\Delta E$. These less significant foci belong to unbound electrons (boundaries for $\Delta E$ can be compared with table \ref{tab1:velocity}) and the apparent period of their oscilation corresponds to the smaller peak visible in figure \ref{fig:sub2}. We can also plot the corresponding electron spectra evolutions in real space (see figure \ref{obr04:coordspectra}), where we observe the same periodical focusing and blurring of the electron spectral density. In the asynchronous case (figure \ref{obr04:coordspectra} on the right), we can see displacement of spectral density peaks as well as the unbound electrons travelling across the central period.

\begin{figure}[t!]\centering
\includegraphics[scale=0.47]{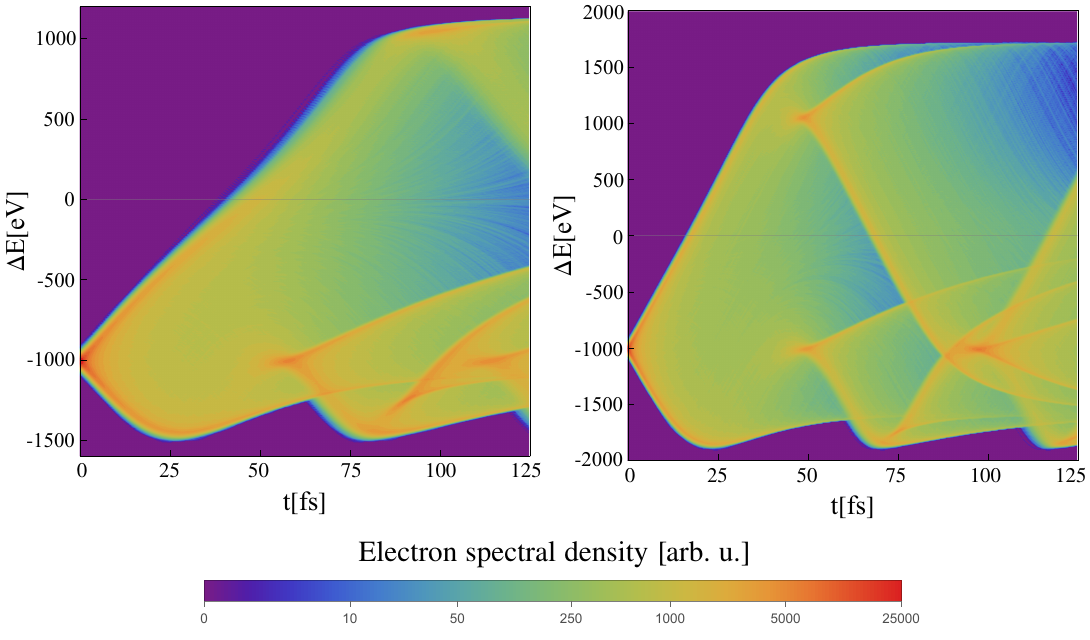}
\caption{\justifying Two cases of electron spectral density evolutions in the asynchronous strong interaction regime, both with initial offset of kinetic energy $\Delta E_0$ with normal distribution characterized by $\mu_E^{(a)}=-1\,\rm keV$ and \mbox{${\Delta E}_{\text{FWHM}}=50\,\rm eV$.} Figure on the left corresponds to $A=30\,\rm eV$ which binds $\approx 25\,\%$ of electrons and figure on the right is obtained by choosing $A=67\,\rm eV$ which binds $\approx 57\,\%$ of electrons.}
\label{obr05:stronginterspectra}
\end{figure}

In order to demonstrate the aspects of strong interaction, we will visualize evolution of electron spectra in this regime as well. The initial offset of kinetic energy (relative to the synchronous case) is assumed to be significantly greater and so will be the subsequent modulation of electron kinetic energy (it can be determined from table \ref{tab1:velocity}). One of the objectives is to illustrate the impact of unsufficient potential amplitude on the resulting evolution. To achieve that, we will again choose normal distribution for $\Delta E_0$ as the initial electron energy spectrum, but with greater offset parameters $\mu_E^{(a)}=-1\,\rm keV$ and ${\Delta E}_{\text{FWHM}}=50\,\rm eV$. Figure \ref{obr05:stronginterspectra} depicts evolution of such a spectrum in two cases, one where the amplitude $A=30\, \rm eV$ corresponds to only $\approx 25\,\%$ of bound electrons (left) and is therefore unsufficient to form alternating spectral density foci as were seen in figure \ref{obr03b:energyspectrumasynch}. In the second case (seen in figure \ref{obr05:stronginterspectra} on the right), we take $A=67\, \rm eV$ which binds $\approx 57\,\%$ of electrons and that is visibly a sufficient bound electron fraction for the emergence of mentioned alternating spectral density peaks. Moreover, one finds that for this particual amplitude and sets of initial conditions, the peak heights in oscillation period distributions, analogous to figure \ref{fig:sub2}, are equal and also the period corresponding to the majority of unbound electrons is exactly half compared to the bound electrons. This means that after the first main unbound period, the foci belonging to both types of electrons are of equal height and temporally aligned, thus serving as an effective beam splitter.

\subsection{Extension to non-parallel scattering}
\begin{figure}[t!]\centering
\includegraphics[scale=0.4]{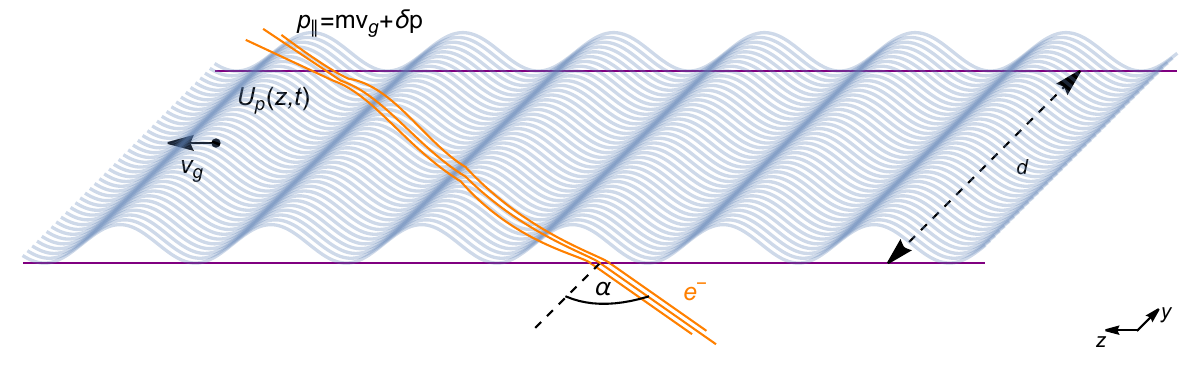}
\caption{\justifying Simple diagram of a collimated electron beam that is being scattered by a ponderomotive potential, moving with group velocity $v_g$ along z-axis, at an incident angle $\alpha$. Duration of the interaction is determined by the width $d$ of the potential's profile in y-direction. }
\label{obr06:neparaleldiagram}
\end{figure}

Previous results are also very useful when we try to obtain an evolution of electron spectra undergoing non-parallel scattering on a moving periodical potential. The situation that will be discussed is depicted on figure \ref{obr06:neparaleldiagram}. An important difference is that the incident angle $\alpha$, potential's group velocity $v_g$ and electron's initial kinetic energy $E_0=\frac{1}{2}m(v_\parallel^2+v_\perp^2)$ must be matched in such a way that for $v_\parallel=v_g+\delta v$, it holds that $\delta v \lll c$ in accordance with the calculations in section \ref{subsec11}.

We use the simplest \footnote{A more realistic potential profile would be to use a Gaussian envelope in the perpendicular direction, which would not be difficult numerically, but we wish to illustrate solely the evolution of parallel momentum spectra.} form of non-parallel scattering, where the potential's profile in perpendicular direction is assumed to be rectangular, therefore the velocity component $v_\perp$ remains unchanged. This fact leads to a linear relation $\Delta t=\frac{d}{\cos \alpha}\sqrt{\frac{m}{2E_0}}$, between the interaction duration $\Delta t$ and the potential's width $d$, so it enables us to visualize the spectral evolution depending on the width of the potential.

\begin{figure}[t!]\centering
\includegraphics[scale=0.46]{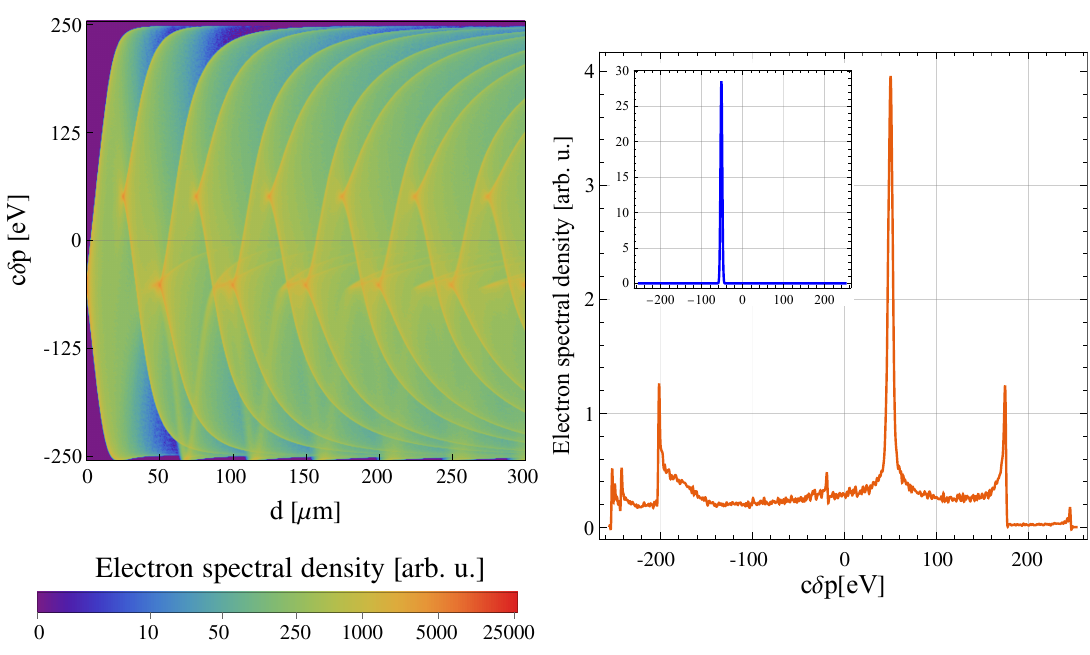}
\caption{\justifying Evolution of the electron parallel momentum spectra $\delta p$ undergoing non-parallel scattering depending on width $d$ of the potential's profile (left) and an electron spectrum corresponding to the narrowest part at $d\approx75\,\rm\mu \rm m$ (right) with the inset representing initial conditions. Results were acquired for incident angle $\alpha=\pi /6$, potential with $A=30\,\rm meV$, $v_g=0.1001\,c$, $\lambda=206\,\rm nm$ and initial electron distribution with Gaussian energy profile $\mu^{(a)}_{E_0}=\frac{m}{2}(0.2c)^2$, ${\Delta E_0}_{\text{FWHM}}=2\,\rm eV$ and constant spatial profile.}
\label{obr07:pplot}

\end{figure}

The subsequent simulation of  electron spectral evolution for our potential (given by $A,\, v_g$ and $\lambda$) and a given incident angle $\alpha$ uses identical expressions for trajectories as in previous simulations. The only difference is that we have to calculate the initial parallel electron velocities in the potential's restframe from their total kinetic energies, giving the result $\delta v=\sin \alpha \sqrt{\frac{2}{m}E_0}-v_g$. We visualized the evolution in momentum space, specifically a dependance of the $\delta p=p_\parallel-mv_g$ spectrum on potential's width $d$. The narrow part of spectrum visible on figure \ref{obr07:pplot} clearly demonstrates the formation of a sharp peak corresponding to opposite sign of the initial $\delta p$ as well as smaller peaks that belong either to the unbound electrons or defocusing of the previous bound peak.

\subsection{Dynamics based on discovered wavefunctions}
Since we have already studied limitations of the interaction, depiction of probability densities stemming from discovered wavefunctions can utilize the suitable physical parameters used when discussing classical dynamics. We will, in both parabolic approximation and free electron evolution in full potential, focus on the weak interaction regime, analyzed on figures \ref{obr01:weakcapturedelectronfraction}, \ref{obr02:periods}, \ref{obr03:energyspectra} and \ref{obr04:coordspectra}. The main goal is to demonstrate that the periodical alternation of sharp peaks over positive and negative values of initial $\Delta E$ takes place in the quantum picture as well. The visualized quantity will be, in correspondence with classical simulations, $|\Psi(t,\,p(\Delta E))|^2$ that is the probability density of an electron having momentum $p$ in the potential's restframe parametrized \footnote{$p(\Delta E)=mv_g\left(\sqrt{1+\frac{2\Delta E}{mv_g^2}}-1\right)$} by the offset of kinetic energy $\Delta E$. This parametrization is unambiguous, since we have assumed relation $p\ll mv_g$ to hold throughout all of the calculations.

First we will address the case of the potential being approximated by a parabola, the applicability of which is limited to wavefunctions that are localized close to the potential's minimum and are unaffected by adjacent spatial periods. A pair of simple such wavefunctions

\begin{align}
\Phi_0(p)&=\frac{1}{\sqrt[4]{2\pi \sigma_p^2}}e^{-\frac{(p-\mu_p)^2}{4\sigma_p^2}}\,,\label{eq037:gaussian}\\
\phi_0^{(n)}(q)&=e^{\frac{i\mu_p q}{\hbar}}\sqrt{\frac{2^\frac{1}{2n}}{\frac{\lambda}{3}\Gamma\left(\frac{2n+1}{2n}\right)}}e^{-\left(\frac{6q}{\lambda}\right)^{2n}},\:n\in\mathbb{N}\,, \label{eq038:supergaussian}
\end{align}

\begin{figure}[b!]\centering
\includegraphics[scale=0.48]{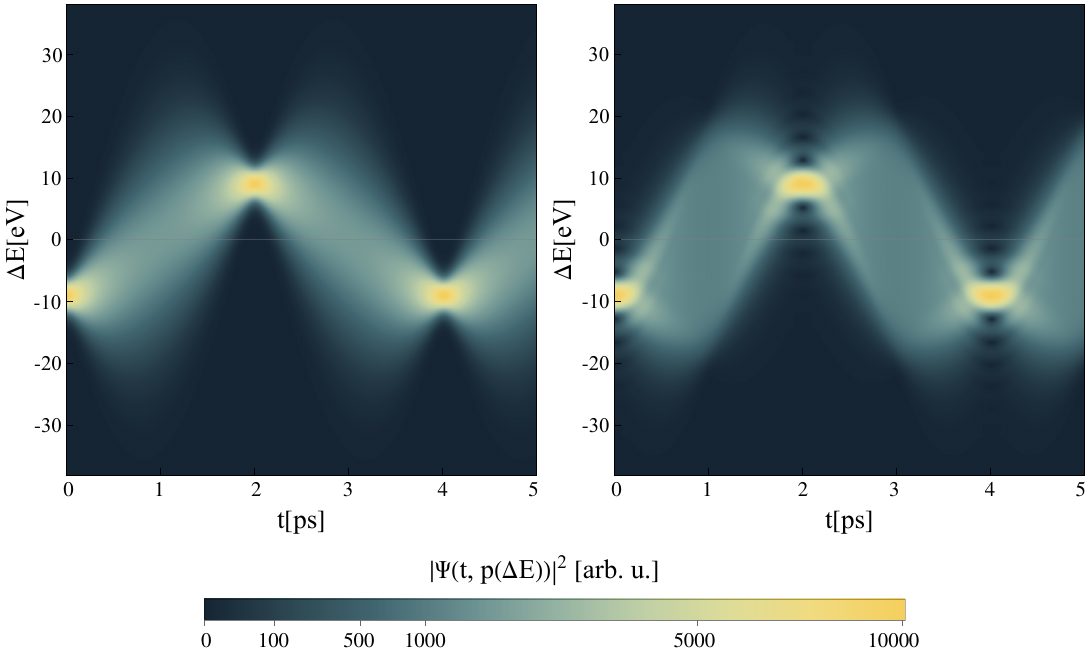}
\caption{\justifying Evolutions of electron probability densities in momentum representation, parametrized by the offset of kinetic energy $\Delta E$ within parabolic approximation of the potential. Gaussian wavepacket evolution (left) corresponds to function \eqref{eq037:gaussian} with $\Delta q_{\text{FWHM}}=\lambda/6$ and super-Gaussian wavepacket evolution (right) corresponds to function \eqref{eq038:supergaussian} of order $n=4$. Both evolutions use potential's spatial period $\lambda=206\,\rm nm$, amplitude $A=30\,\rm meV$ and initial wavefunctions are centered around $\mu_p$ corresponding to initial offset of kinetic energy $\Delta E_0=-9\,\rm eV$. }
\label{obr08:lhoplot}
\end{figure}

\begin{figure}[b!]\centering
\includegraphics[scale=0.8]{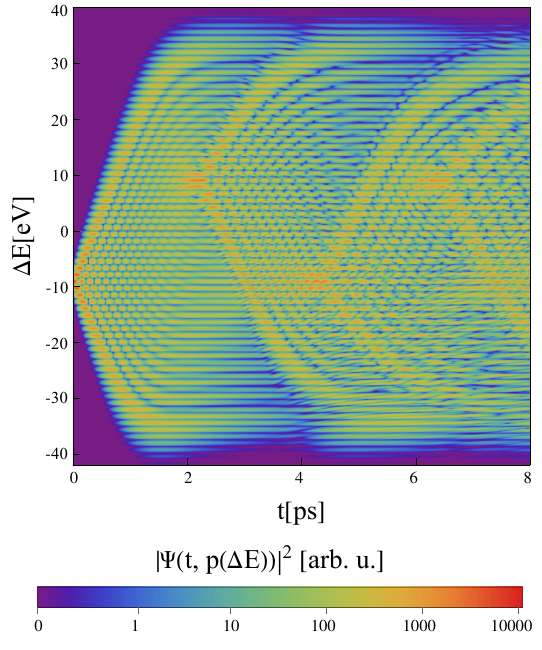}
\caption{\justifying Evolution of electron probability density in momentum representation, parametrized by the offset of kinetic energy $\Delta E$ within full periodical potential. Initial electron wavefunction consisted of discrete momentum eigenstates sampled from a Gaussian distribution of $\Delta E_0$ with $\mu_E^{(a)}=-9\,\rm eV$ and ${\Delta E}_{\text{FWHM}}=0.5\,\rm eV$ adjusted for momenta. Individual eigenstates were then blurred according to the sampling distribution in order to better approximate the original wavepacket.  The evolution uses potential's spatial period $\lambda=206\,\rm nm$ and amplitude $A=30\,\rm meV$.}
\label{obr09:blochplot}

\end{figure}

\noindent will be used to perform numerical simulations. Function \eqref{eq037:gaussian} is the standard Gaussian wavepacket with mean momentum $\mu_p$ and variance $\sigma_p^2$. Its Fourier transform is also a Gaussian wavepacket with mean coordinate $\mu_q=0$ and variance $\sigma_q^2$ such that $\sigma_q^2\sigma_p^2=\hbar^2/4$. To satisfy the condition for localization near the potential's minimum, we will choose $\sigma_q$ corresponding to $\Delta q_{\text{FWHM}}=\lambda/6$. The function \eqref{eq038:supergaussian} is a normalized super-Gaussian distribution of order $n$, the profile of which is also centered in $q=0$ and with slightly wider localization, its corresponding momentum representation is centered around $\mu_p$. The advantage of super-Gaussian distribution lies in its flatter profile that better resembles the constant distribution of initial positions used in classical cases \footnote{Of course that the momentum representation is no longer a simple Gaussian wavepacket due to uncertainty relations.}, but its function is otherwise mainly illustrative, since coherent super-Gaussian wavepackets have no reported way of being generated. Figure \ref{obr08:lhoplot} depicts the numerically acquired evolutions of these wavefunctions within the parabolic approximation of our potential. Their main feature is the above mentioned formation of alternating sharp peaks in associated electron probability densities. Specifically for the evolution of a simple Gaussian wavepacket (figure \ref{obr08:lhoplot} on the left), we can easily make a comparison with non-linear classical trajectories, for which we can choose initial conditions $(z_0,\,\delta v)$ by sampling from Gaussian distributions $|\Phi_0(m\delta v)|^2$ and its Fourier transform $|\phi_0(q)|^2$. We would find a very good quantitative agreement with slight differences in the region of higher $|\Delta E|$ due to non-linearity of the classical trajectories.

Finally, we will discuss electron spectral evolution in full periodical potential as it was described in section \ref{sec22}. Analytical approach that was used when solving the stationary Schrödinger equation does allow us to evolve any wave function composed of a finite number of plane waves, which can be viewed as an approximation

\begin{equation}
\phi_0(q)=\frac{1}{\sqrt{2\pi\hbar}}\int \Phi_0(\hbar k)e^{-ikq}d(\hbar k)\approx\sum_j \omega_j e^{ik_jq}\,,
\end{equation}

\noindent where the weights $\omega_j$ corresponding to plane wave with momentum $p_j=\hbar k_j$ stem from the original wavepacket $\Phi_0(p_j)$. Linearity of the Schrödinger equation allows us to find a solution for each plane wave described by $k_j$ and use the weighted superposition of these solutions 
as an approximation for the whole wavepacket evolution.

This approach is also not that demanding numerically. For each momentum eigenstate $e^{ik_jq}$ we construct the tridiagonal Hamiltonian, analytically expressed by relation \eqref{eq034:matrixelem} and solve its eigenvalue problem, for which many efficient algorithms are known, yielding a set of eigenvectors $\{c_{nj}\}_{j=-j_{min}}^{j=j_{max}}$ and eigenvalues $E_n$. A simple numerical examination also concludes that the convergence rate of coefficients $c_{nj}$ is many orders of magnitude faster than the divergence of eigenvalues with increasing matrix size so that rapidly oscillating components are heavily surpressed in the final solution.

The logarithmic scale in figure \ref{obr09:blochplot} clearly shows that even with a discrete initial momentum eigenstates, the formation of alternating sharp peaks in electron probability densities resumes even with a full periodical potential in the quantum picture. The total broadening of the energy spectrum also corresponds to the kinetic energy modulation limits calculated in the classical analysis. Notice also that the subsequent peak over negative $\Delta E$ is accompanied by a smaller increment on the left, which is the result of an unbound component of the wavefunction that is in full analogy with the classical evolution depicted on figure \ref{obr03b:energyspectrumasynch}. A closer look also shows a slightly shifted period of oscillation, which matches the classical predictions (seen in figures \ref{obr02:periods}).

In the case of plotting the evolution of a single momentum eigenlevel in a potential whose strength corresponds to equal intensities of bound and unbound components, we would also discern two simultaneous peaks in the spectrum in the first focusing region, as in figure \ref{obr05:stronginterspectra} on the right. However, their height would be greatly blurred by quantum interference due to the infinite spatial width of the electron packet with respect to the spatial period of the potential.

\subsection{Free electron interaction with pulsed laser beams}
In this chapter we will compare the classical results obtained in chapter \ref{chapter3.1} with more realistic case of the electrons interacting with pulsed laser beams with Gaussian spatial and time envelopes. The numerical simulations are performed using General Particle Tracer (GPT) code with embedded fifth order Runge-Kutta algorithm solving the relativistic equation of motion for each electron:
\begin{equation}
\frac{d\left(\gamma m_0{\bm v}\right)}{dt}=q\left({\bm E}+{\bm v}\times{\bm B}\right).
\end{equation}
\noindent Here $\bm E$ and $\bm B$ are electric and magnetic fields of two optical pulses with Gaussian envelopes, $m_0$ is electron rest mass and $\gamma=1/\sqrt{1-\beta^2}$ is Lorentz factor with $\beta$ being the ratio of electron speed $\left| \bm v \right|$ to the speed of light $c$. For electric and magnetic fields of pulsed laser beams we use the paraxial approximation (which is sufficient as we do not assume strongly focused beams). Electric field of a linearly polarized laser pulse propagating in $z$ direction is defined as
\begin{eqnarray}
E_y(r,\,z,\,t)&=&E_0\frac{w_0}{w\left(z\right)}e^{{-2\ln{2}\left(\frac{t-\frac{z}{c}}{t_\mathrm{FWHM}}\right)}^2}\nonumber \\
 & &\times e^{{-\left(\frac{r}{w(z)}\right)}^2}\sin{\left[\phi+\rho(z)\right],}
\end{eqnarray}
\noindent where $E_0$ is the amplitude of electric field, $r=\sqrt{x^2+y^2}$,  $w_0$ is $1/e^2$ beam waist radius, $w(z)=w_0\sqrt{1+\left(z/z_\mathrm{R} \right)^2}$ is beam radius at distance $z$ from the beam waist, $z_\mathrm{R}=\pi w_0^2/\lambda$ is Rayleigh length with $\lambda$ being the central wavelength of the pulse, $t_\mathrm{FWHM}$ is the duration of the pulse defined as full width at half maximum of its intensity distribution, $\phi=\omega t-kz-\frac{kr^2}{2R(z)}$  with  $R(z)=z\left[1+\left(z_\mathrm{R}/z\right)^2\right]$ being the radius of curvature of the wavefront, $k=2\pi/\lambda$  is the wavenumber, $\omega=2\pi c/\lambda$ is the central angular frequency and $\rho(z)=\mathrm{arctan}\left(z/z_\mathrm{R}\right)$ is Gouy phase shift.

For numerical simulations we set the parameters of the optical pulses to have the central wavelengths of $\lambda_1=343\rm\: nm$  and $\lambda_2=515\rm\: nm$ (second and third harmonic frequencies of the fundamental wavelength 1030 nm), electric field amplitudes of $E_0=1.85 \cdot {10}^9\rm\: V.m^{-1}$, beam waist radii $w_0=100\rm\: \mu m$, and angles of incidence with respect to the direction of the electron trajectory $\alpha=90^{\circ}$. The value of electric field amplitude $E_0$ is chosen to generate harmonically modulated ponderomotive potential with amplitude of $A=30\rm\: meV$ used in earlier simulations in chapter \ref{chapter3.1} as
\begin{equation}
E_0=\sqrt{\frac{Am_0\omega_1\omega_2}{e^2},}
\end{equation}
\noindent where e is elementary charge.

Pulse durations $t_\mathrm{FWHM}$ are set to be changing from 0 to 10 ps, which correspond to the pulse energies up to 0.76 mJ. The effective interaction time $t_\mathrm{eff}$ of electrons with the optical travelling wave created by two counterpropagating pulses is determined from the temporal integral of the Gaussian function as
\begin{equation}
t_\mathrm{eff}=t_\mathrm{FWHM}\sqrt{\frac{\pi}{4\ln{2}\left(1+\beta^2\right)}}.
\end{equation}
\noindent In the synchronous regime, $\beta=0.2$.

\begin{figure}[t!]\centering
\includegraphics[scale=0.25]{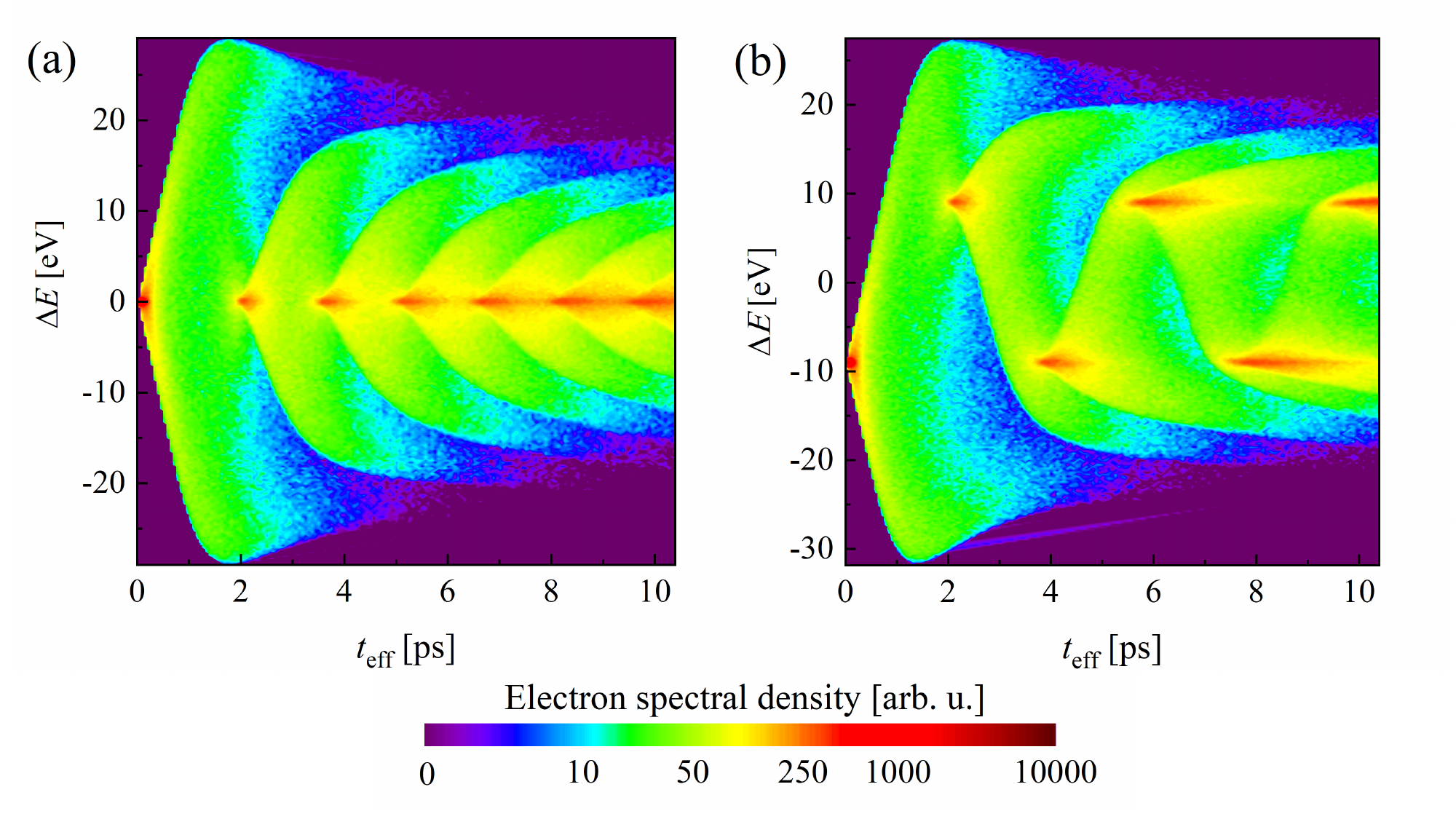}
\caption{\justifying Interaction of electrons with Gaussian envelopes: evolution of electron spectral densities in energy space for (a) synchronous and (b) asynchronous case of the interaction with $\mu_E^{(a)}=-9\rm\: eV$.}
\label{obr:gauss}
\end{figure}

\begin{figure}[t!]\centering
\includegraphics[scale=0.25]{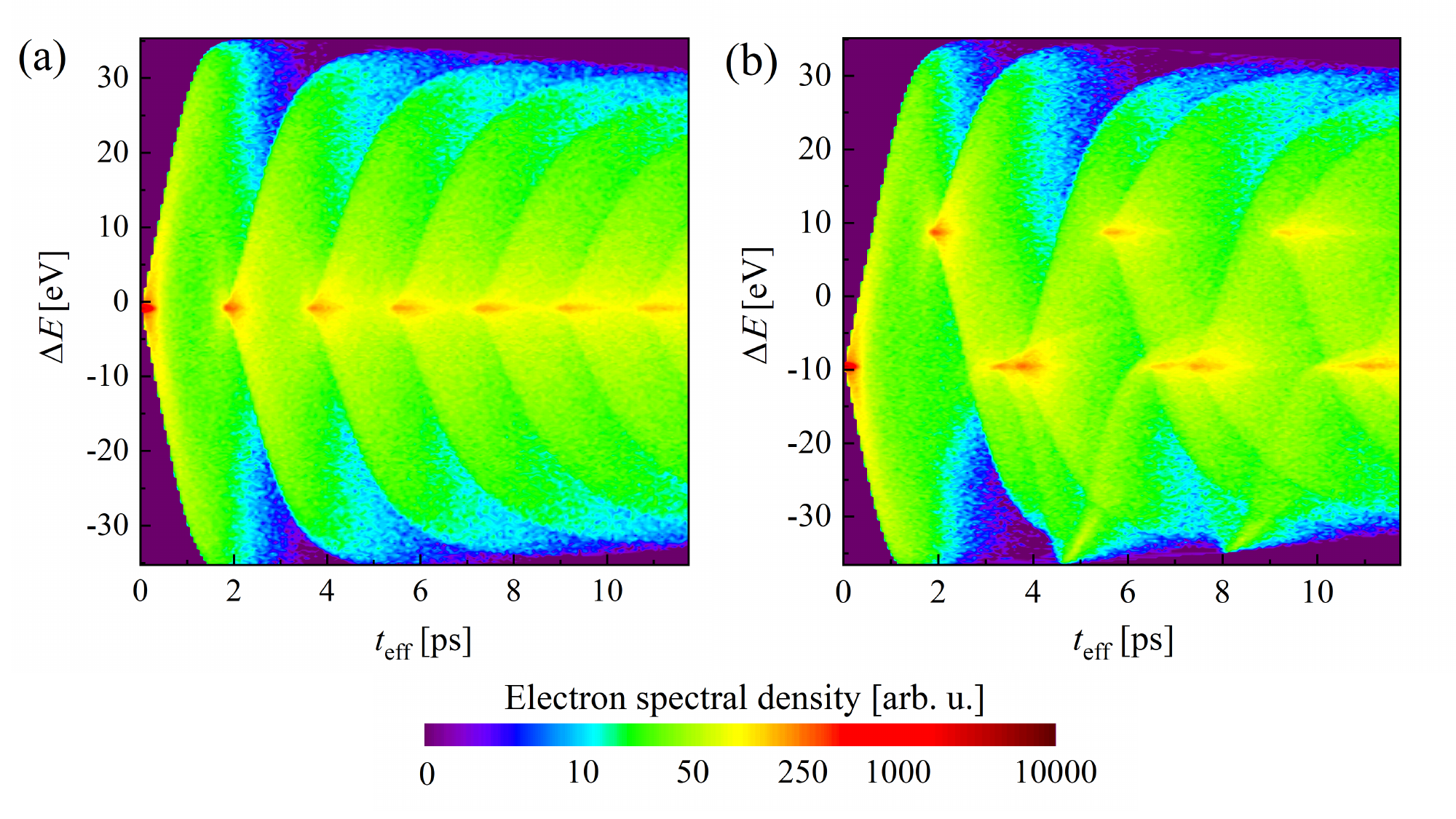}
\caption{\justifying Interaction of electrons with super-Gaussian envelopes: evolution of electron spectral densities in energy space for (a) synchronous and (b) asynchronous case of the interaction with $\mu_E^{(a)}=-9\rm\: eV$.}
\label{obr:supergauss}
\end{figure}

We assume that the initial offsets of electron kinetic energy $\Delta E$ are sampled from a Gaussian distribution with mean value $\mu_E^{(s)}=0\rm\: eV$ for synchronous case, $\mu_E^{(a)}=-9\rm\: eV$ for asynchronous case and ${\Delta E}_{\text{FWHM}}=0.5\rm\: eV$ for both cases, again in accordance with the parameters of the previous simulations. The initial distribution of electron positions is set to be Gaussian with the transverse width of $\sigma_{xy}=20\rm\: \mu m$ and the longitudinal width of $\sigma_z=30\rm\: \mu m$.
The results of numerical simulations for ${10}^4$ electrons are shown in figure \ref{obr:gauss}, where we plot the temporal evolution of electron energy spectra in the ponderomotive potential generated by two optical pulses described above for (a) synchronous and (b) asynchronous case of the interaction. Similar to figure \ref{obr03:energyspectra} in chapter \ref{chapter3.1}, we see periodic focusing and blurring of the spectral signal with even spectral peaks being shifted from the initial electron energy for asynchronous regime. However, unlike in figure \ref{obr03:energyspectra} we now see that with increasing interaction time $t_\mathrm{eff}$ the maximum kinetic energy difference is decreasing, periodic foci are gradually spreading in time and the distance of the foci is not equidistant and changes with the size of the asynchronicity. Furthermore, fast oscillating foci belonging to unbound electrons are missing.

We now consider super-Gaussian pulses to show that the above mentioned differences between figure \ref{obr03:energyspectra} and \ref{obr:gauss} are given mainly due to the changing magnitude of the electric field that the propagating electron experiences while interacting with the Gaussian pulses. We define the electric field of super-Gaussian optical pulses of order 10 as
\begin{eqnarray}
E_y(r,\,z,\,t)&=&E_0\frac{w_0}{w\left(z\right)}e^{{-\ln{2}\left[{2\left(\frac{t-\frac{z}{c}}{t_\mathrm{FWHM}}\right)}^2\right]}^{10}}\nonumber\\
&&\times
e^{{-\left(\frac{r}{w(z)}\right)}^2}\sin{\left[\phi+\rho(z)\right]}
\end{eqnarray}
\noindent and run other numerical simulations with the same set of parameters as in previous case with Gaussian pulses. Note that for super-Gaussian pulses $t_\mathrm{eff}=t_\mathrm{FWHM}$. The results of simulations for ${10}^4$ electrons are shown in figure \ref{obr:supergauss} and closely corresponds to the results obtained in chapter \ref{chapter3.1}.

Figure \ref{obr:gauss} shows that the finite length of laser pulses and Gaussian profile of their envelopes lead to the gradual temporal spread of periodic foci during the evolution of electron spectral densities and at the same time to the gradual reduction of the electron spectral width. An interesting phenomenon occurs when the interaction time is sufficiently long such that subsequent periods of oscillations overlap. In this regime, the scheme could effectively act as an electron beam splitter. To demonstrate such effect, we change the geometry of the experiment. In previous simulations we assumed inelastic interaction leading to the longitudinal momentum transfer. Alternatively we can assume different case with perpendicular geometry where the electrons are being scattered by an optical standing wave created by two counterpropagating laser pulses of the same frequency (as in the original Kapitza-Dirac experiment \cite{kapitza1933reflection}). Now the interaction is elastic and leads to transverse momentum transfer and thus to undirectional acceleration of electrons in the transverse direction. Asynchronicity is in this case introduced by the nonzero angle of incidence $\alpha$ (deviation from the perpendicular geometry of laser beams and electrons’ trajectory).

In the simulation we set all electrons to speed $v=0.01c$. The initial distribution of electron positions is again Gaussian with distribution widths $\sigma_{xy}=20\rm\: \mu m$ and $\sigma_z=10\rm\: \mu m$. Both laser pulses are set to central wavelength $\lambda=1030\rm\: nm$, electric field amplitude $E_0=8 \cdot {10}^9\rm\: V^{-1}$ and beam waist radius $w_0=100\rm\: \mu m$. The value of the electric field amplitude $E_0$ now corresponds to the ponderomotive potential amplitude $A=3.4\rm\: eV$.  Laser pulse durations $t_\mathrm{FWHM}$ are set from 0 to 25 ps, which correspond to the pulse energies up to 35.5 mJ. For this geometry and small angles of incidence $t_\mathrm{eff}=t_\mathrm{FWHM}\sqrt{\frac{\pi}{4\ln{2}}}$.

\begin{figure}[t!]\centering
\includegraphics[scale=0.25]{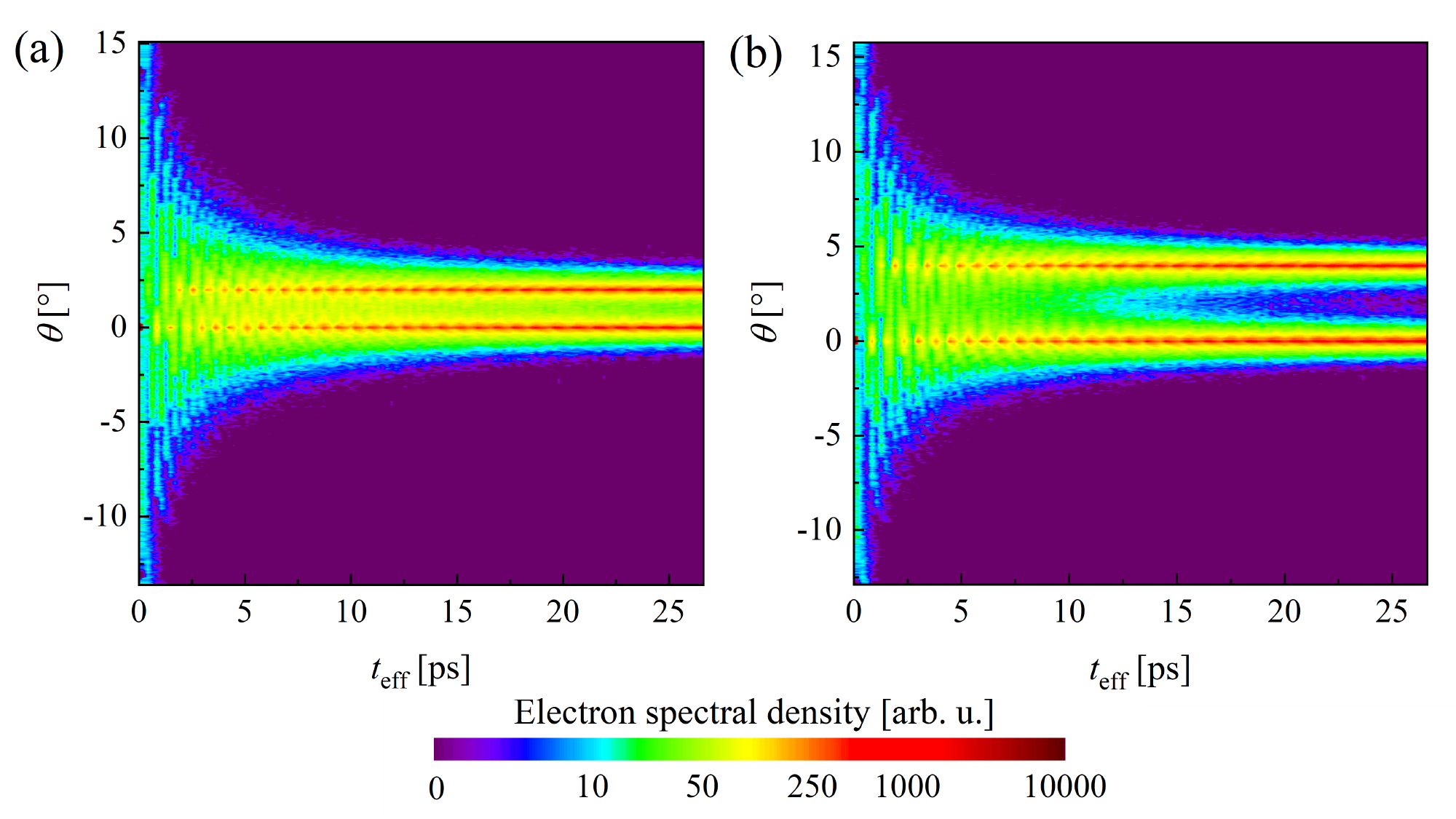}
\caption{\justifying Elastic interaction of electrons with ponderomotive potential: evolution of electrons’ deviation from the original angle of motion $\theta_\mathrm{o}=0^{\circ}$. The asynchronicity is introduced by angle of incidence (a) $\alpha=1^{\circ}$ and (b) $\alpha=2^{\circ}$.}
\label{obr:beamsplitter}
\end{figure}

The results of numerical simulations for ${10}^4$ electrons are shown in figure \ref{obr:beamsplitter}. In plot (a) the asynchronicity is introduced with angle of incidence $\alpha=1^{\circ}$ and in plot (b) with angle $\alpha=2^{\circ}$. Both graphs show the evolution of electron transverse scattering parametrized by the angle $\theta$ which characterizes the deviation from the original direction of motion of electrons. Again we see periodic focusing and blurring, this time in the transverse spatial domain, with periodic foci alternating between the electrons' original angle of motion $\theta_\mathrm{o}=0^{\circ}$ and angle $\theta_\mathrm{d}=2\alpha$. The finite length of pulsed optical beams leads to the gradual temporal blurring of the foci which after some interaction time results in a condition where most of the electrons are either deflected in direction $\theta_\mathrm{d}=2\alpha$ or remain in the original direction $\theta_\mathrm{o}=0^{\circ}$. We refer to this phenomenon as electron beam splitter in spatial domain. An analogous result can be obtained for the inelastic case of interaction, where long interaction times eventually lead to electron beam splitter in the energy domain. We note that this regime of electron scattering is fully classical and does not correspond to Bragg scattering, in which the electron wave interacts with a thick periodic optical standing wave. In our case, the splitting is purely a result of electron oscillations in the individual periods of the optical standing waves, where the potential can be approximated by parabolic function.

\section*{Data availability statement}
All data that support the findings of this study are publicly
available at Zenodo, DOI: \href{https://doi.org/10.5281/zenodo.14918098}{10.5281/zenodo.14918098}.

\section*{Acknowledgments} 

The authors acknowledge funding from the Czech Science
Foundation (project 22-13001K), Charles University
(SVV-2024-260720, PRIMUS/19/SCI/05, GAUK
216222) and the European Union (ERC, eWaveShaper,
101039339). Views and opinions expressed are however
those of the author(s) only and do not necessarily reflect
those of the European Union or the European Research
Council Executive Agency. Neither the European
Union nor the granting authority can be held responsible
for them. This work was supported by TERAFIT
project No. CZ.02.01.01/00/22\_008/0004594 funded by
OP JAK, call Excellent Research.

\appendix
\setcounter{section}{1}
\section*{Appendix A}\label{append}
Let's briefly address a couple of purely mathematical properties of the discovered classical trajectories of unbound and bound electrons, explicitly described by relations \eqref{eq09:unboundtrajec} and \eqref{eq015:boundsolution}. Key properties are encoded in the elliptic functions $am(x,\,m)$ and $sn(x,\,m)$, which are used in the expressions for analytical trajectories.

Trajectories are characterized by functions 
\begin{align}
y_f(x)&=am(x,m)\,,\\
y_b(x)&=\arcsin\left[\sqrt{p}\,sn\left(\frac{x}{\sqrt{p}},\:p\right)\right]\,,
\end{align}
\noindent where index $f$ denotes the unbound trajectory and index $b$ the bound trajectory and for their defining parameters it holds that $m,\,p\in ( 0,\,1)$. Scaling by factor $\sqrt{p}$ in the definition of $y_b (x)$ is not groundless, the factor is present because of a transformation identity (seen in \cite{olver2010nist})
\begin{equation}
sn\left(x,\,\frac{1}{p}\right)=\sqrt{p}\:sn\left(\frac{x}{\sqrt{p}},\,p\right)\,.
\end{equation}
\noindent Function $y_b(x)$ can therefore be very roughly understood as a continuous, real and bound equivalent to the elliptic function  $am\left(x,\,\frac{1}{p}\right)$ for $p\in(0,\,1)$. By looking carefully at the discovered trajectories \eqref{eq09:unboundtrajec} and \eqref{eq015:boundsolution}, we see that they are connected by this exact transformation, if we also realize that the incomplete elliptical integral of the first kind satisfies a similar transformation (seen again in \cite{olver2010nist})
\begin{equation}
F\left(x,\,\frac{1}{k}\right)=\sqrt{k}\,F\left[\arcsin\left(\frac{\sin x}{\sqrt{k}}\right),\,k\right]\,.
\end{equation}

\begin{figure}[t!]\centering
\includegraphics[scale=0.5]{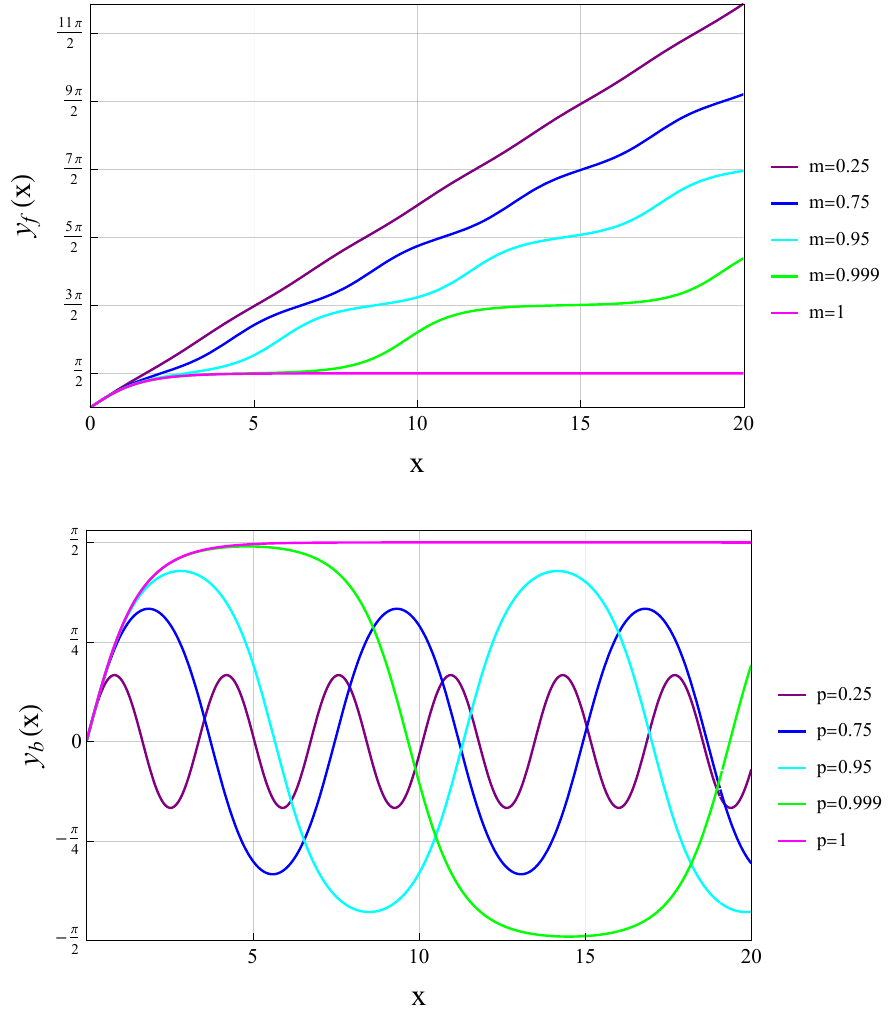}
\caption{\justifying Plots of solutions corresponding to bound $y_b(x)$ and unbound $y_f(x)$ trajectories in a dimensionless problem for several values of the defining parameters $m$ and $p$.}
\label{obr010:exampletrajectories}

\end{figure}

Figures \ref{obr010:exampletrajectories} depict the functions $y_f (x)$ and $y_b (x)$ for several values of parameters $m,\,p\in ( 0,\,1)$, which define their behaviour. We can see from the figures that the functions behave in a more interesting way when the values of $m$ and $p$ are approaching one. In case of $y_f (x)$, these values correspond to a tight overshoot of an electron into the neighbouring spatial period and in case of $y_b (x)$ it means a borderline bounding of the electron. We also see that for the limiting case $m=p=1$ the curves merge and depict the trajectory of an electron which rises to the potential's maximum, where it stops.

An important property, which will come handy when we will analyze trajectories for specific parameters, is periodicity. Function $am(x,\,m)$ is for a given $m\in (0,\,1)$ continuous, monotonous and pseudoperiodic with period $2K(m)$. Therefore it holds that (again seen in \cite{olver2010nist})
\begin{equation}
am(x+2K(m),\,m)=am(x,\,m)+\pi\,.
\end{equation}
\noindent If we take a look at the unbound trajectory \eqref{eq09:unboundtrajec}, we can see that the shift by $\pi$ corresponds exactly to relocation of the electron to the same position within neighbouring spatial period of our potential.

Function $sn(x,m)$ is for $m\in (0,\,1)$ continuous and real with range $\langle -1,\,1 \rangle$. It is also periodic with a period $4K(m)$. Due to $\arcsin$ being monotonous on $\langle -1,\,1 \rangle$, the composite function describing bound trajectory $y_b (x)$ is itself periodical with period $4K(m)$, which is a result that is known for a pendulum without the mechanical energy to make a full turn \cite{belendez2007exact}.

\section*{Appendix B}\label{appended}
In order to outline the derivation and applicability of the assumed potential \eqref{eq01:pot}, we will consider a particle of mass $m$ and charge $q$ with initial ($t=0$) position $\mathbf{r}_0$ and velocity $\mathbf{v}_0$ moving through a bichromatic electromagnetic field
\begin{equation}
\begin{aligned}
\bm E(\bm r,\,t) &= \bm E_1(\bm r,\,t)+\bm E_2(\bm r,\,t) \,,\quad\text{where}  \\
\bm E_i(\bm r,\,t) &= \frac{1}{2}\left(\bm \xi_i(\bm r) e^{i\omega_i t}+\bm \xi_i^\ast(\bm r) e^{-i\omega_i t}\right)\quad\text{for}\; i\in\{1,\,2\} \,.
\end{aligned}
\end{equation}
\noindent We wish to solve its EOM in the initial rest-frame \mbox{$\bm r'\equiv \bm r -\mathbf{v}_0 t-\bm r_0$} and we can therefore write
\begin{equation}
\begin{aligned}
m\ddot{\bm r}'(t)=q\left[\bm E(\bm r'(t)+\bm v_0 t+\bm r_0,\,t)+(\dot{\bm r}'(t)+\bm v_0)\right.\\\left. \times \bm B(\bm r'(t)+\bm v_0 t+\bm r_0,\,t) \right]\,.
\end{aligned}
\end{equation} 
\noindent Solution to the EOM can be approximated as in \cite{bochove1996unified} by assuming small field strengths and expanding \mbox{$\mathbf{r}'(t)=\bm R^{(1)}(t)+\bm R^{(2)}(t)+\dots{}\,,$} where $\bm R^{(i)}(t)$ is proportional to the $i$-th power of the fields. By also expanding the fields spatial profile in the particle's initial rest-frame displacement, we get the first two equations

\begin{equation}
\begin{aligned}
m&\frac{d^2\bm R^{(1)} (t)}{dt^2}=q\left[\bm E (\bm r_0+\bm v_0t,\,t)+\bm v_0\times \bm B (\bm r_0+\bm v_0t,\,t)\right]\,, \\
m&\frac{d^2\bm R^{(2)} (t)}{dt^2}=q\left\{\bm R^{(1)} (t)\boldsymbol{\cdot}\bm\nabla\bm E(\bm r_0+\bm v_0t,\,t)+\frac{d\bm R^{(1)} (t)}{dt}\right.\\ &\times\left.\bm B(\bm r_0+\bm v_0t,\,t)+ \bm v_0\times\left[\bm R^{(1)} (t)\boldsymbol{\cdot}\bm\nabla\bm B(\bm r_0+\bm v_0t,\,t)\right] \right\} \,.
\end{aligned}
\end{equation}

This system can be analytically solved by assuming a 1D geometry consisting of two counterpropagating plane-waves with matching amplitudes and different frequencies, which move along the particle's trajectory. In this simplification, the $\bm R^{(1)} (t)$ component of motion corresponds to oscillations parallel with the electric field polarization direction. The second-order force is therefore represented only by
\begin{equation}
\bm F^{(2)}(\bm r_0,\,t)=q\frac{d\bm R^{(1)}(t)}{dt}\times \bm B (\bm r_0+\bm v_0t,\,t)
\end{equation} 
\noindent and it acts along the particle's trajectory. By assuming that the particle initially moved with velocity only slightly differing from the group velocity of the composite wave $v_0\approx v_g = c\,\frac{\omega_1-\omega_2}{\omega_1+\omega_2}$, we can average this force over time to cancel terms oscillating with $\omega_i$, return to the original laboratory reference frame and integrate to obtain the corresponding potential

\begin{equation}
\begin{aligned}
U_p(z,\,t)=\frac{q^2 E_0^2}{2m\omega_1 \omega_2}\cos\Bigr(\frac{z(\omega_1+\omega_2)}{c}-(\omega_1-\omega_2)t\\+\varphi_2-\varphi_1\Bigr)+const.\;,
\end{aligned}
\end{equation}
\noindent where we assumed that the particle moves along the $z$-axis. We also denoted $E_0$ the amplitude of the electric field components and $\varphi_i$ their constant phases. 

In the initial laboratory frame, we therefore obtain (by choosing appropriate constant and transforming away the constant phase) the moving ponderomotive potential experienced by a charged particle drifting slowly within the potential's rest frame
\begin{equation}
U_p(z,\,t)=\frac{A}{2}\left(1-\cos\frac{2\pi (v_g t-z)}{\lambda}\right)\,,
\end{equation}
\noindent where we denoted
\[
A\equiv\frac{q^2 E_0^2}{m\omega_1\omega_2}\,, \quad v_g\equiv c\,\frac{\omega_1-\omega_2}{\omega_1+\omega_2}\,,\quad \lambda\equiv\frac{2\pi c}{\omega_1+\omega_2}\,.
\]
\noindent The use of this potential is justified only in the case of a persisting slow drift relative to the potential's velocity. The restriction to a 1D potential is justified only in the case of a narrowly focused transverse electron beam profile compared to the transverse spatial modulation of the laser pulses.


%

\end{document}